\documentclass[aps,prx,preprint,bibliography,superscriptaddress]{revtex4-1}
\usepackage[utf8]{inputenc}
\usepackage{amsmath}

\usepackage{physics}
\usepackage{mathtools}
\usepackage{amsfonts}
\usepackage{amssymb}
\usepackage{graphicx}
\usepackage{wrapfig}
\usepackage{textcomp}
\usepackage{color}
\usepackage[dvipsnames]{xcolor}
\usepackage{comment}

\begin{document}

\title{Non-Hermitian Topological Phases: Principles and Prospects}

\author{Ayan Banerjee}
\affiliation{Solid State and Structural Chemistry Unit, Indian Institute of Science, Bangalore 560012, India}
\author{Ronika Sarkar} 
\affiliation{Department of Physics, Indian Institute of Science, Bangalore 560012, India}
\affiliation{Solid State and Structural Chemistry Unit, Indian Institute of Science, Bangalore 560012, India}
\author{Soumi Dey}
\affiliation{Solid State and Structural Chemistry Unit, Indian Institute of Science, Bangalore 560012, India}
\author{Awadhesh Narayan}
\email{awadhesh@iisc.ac.in}
\affiliation{Solid State and Structural Chemistry Unit, Indian Institute of Science, Bangalore 560012, India}

\date{\today}

\begin{abstract}
The synergy between non-Hermitian concepts and topological ideas have led to very fruitful activity in the recent years. Their interplay has resulted in a wide variety of new non-Hermitian topological phenomena being discovered. In this review, we present the key principles underpinning the topological features of non-Hermitian phases. Using paradigmatic models -- Hatano-Helson, non-Hermitian Su-Schrieffer-Heeger and non-Hermitian Chern insulator -- we illustrate the central features of non-Hermitian topological systems, including exceptional points, complex energy gaps and non-Hermitian symmetry classification. We discuss the non-Hermitian skin effect and the notion of the generalized Brillouin zone, which allows restoring the bulk-boundary correspondence. Using concrete examples, we examine the role of disorder, describe the Floquet engineering, present the linear response framework, and analyze the Hall transport properties of non-Hermitian topological systems. We also survey the rapidly growing experimental advances in this field. Finally, we end by highlighting possible directions which, in our view, may be promising for explorations in the near future.
\end{abstract}

\maketitle

\tableofcontents

\section{Introduction}

Hermiticity is a central pillar of quantum mechanics~\cite{dirac1981principles}, which dictates that the observables be represented by Hermitian or self-adjoint \footnote{Note that the definition for the adjoint of an operator $\mathrm{\it{O}}$, denoted $\mathrm{\it{O^*}}$ is, for a given notion of an inner product and vectors $\mathrm{v},\mathrm{w}, (\mathrm{\it{O}}\mathrm{v},\mathrm{w})=(\mathrm{v},\mathrm{\it{O^*}}\mathrm{w})$, which can be written as $\mathrm{\it{O^*}}= G^{-1}\mathrm{\it{O^{\dagger}}}G$, where $G$ denotes the Gram matrix of  the corresponding inner product. An operator is said to be self-adjoint if $\mathrm{\it{O^*}}=\mathrm{\it{O}}$. In Hermitian quantum mechanics, $G=1$, the identity matrix, and hence the notion of an operator being self-adjoint is equivalent to $\mathrm{\it{O}}=\mathrm{\it{O^{\dagger}}}$. In non-Hermitian quantum mechanics, $G$ could be chosen differently, and the notion of self-adjointness changes~\cite{ju2019non}.} operators. This leads to conservation of probability and real eigenvalues of such operators. However, since the early days of quantum mechanics it was noted that many systems exhibit a lack of probability conservation due to exchange of particles or energy with their surroundings~\cite{moiseyev2011non}. In the recent decades, pioneering work by Bender and co-workers radically transformed the understanding of non-Hermitian Hamiltonians~\cite{bender1998real,bender2007making}. This field has today burgeoned into the exciting forefront of non-Hermitian systems. Interestingly, not only quantum, but also a variety of classical systems can be non-Hermitian in nature. It has been realized that non-Hermitian systems exhibit many properties that are unique to them and do not have any analogs in Hermitian systems. 

The last decades have witnessed a growing interest in ideas from topology in physics, beginning with the quantum Hall effect~\cite{stone1992quantum}. Remarkable connections to topology have culminated in the celebrated topological band theory, which is at play in topological insulators, topological semimetals, and topological superconductors~\cite{hasan2010colloquium,qi2011topological,armitage2018weyl,lv2021experimental}. A cornerstone in the analysis of topological phases has been the bulk-boundary correspondence -- bulk topology is reflected in the existence of symmetry-protected boundary states. Often such topological phases also harbor protected band crossings, which lead to exotic experimental signatures. Moreover, an astonishing range of experimental platforms have been recognized to host these exciting phenomena, ranging from condensed matter materials~\cite{yan2012topological}, ultra-cold atoms~\cite{cooper2019topological,zhang2018topological}, photonic~\cite{lu2014topological,ozawa2019topological} and phononic~\cite{liu2020topological1} systems, to name just a few.

In recent years, the synergy between non-Hermitian concepts and topological ideas has led to extremely fruitful and rapid activity. Their union has resulted in a wide variety of new non-Hermitian phenomena being discovered and reinterpreted within the framework of topology. We note here that there are several excellent reviews on complementary aspects of non-Hermitian systems~\cite{martinez2018topological,ghatak2019new,torres2019perspective,ashida2020non,bergholtz2021exceptional,zhang2022review,ding2022non}. In this review, we focus on the key concepts and ideas underpinning topological features of non-Hermitian systems, and summarize the latest developments in this rapidly evolving field. We start with a discussion of exceptional points, which are ubiquitous in non-Hermitian topological phases and can be thought of as an intriguing generalization of Hermitian band degeneracies. We next present paradigmatic models which have played a vital role as playgrounds for discovering many of the salient features of non-Hermitian topological phases. These include the Hatano-Nelson and non-Hermitian Su-Schrieffer-Heeger models in one dimension, the non-Hermitian Chern insulator model in two dimensions and non-Hermitian topological semimetals in three dimensions. We thoroughly discuss their spectral topology and use them to illustrate many of the important features of non-Hermitian topological systems. We illustrate the Floquet engineering of non-Hermitian phases and highlight the recent advances made in this direction. We then present a detailed discussion of the non-Hermitian symmetry classes which generalize the celebrated Altland-Zirnbauer classification of Hermitian systems. We illustrate these ideas using the non-Hermitian Su-Schrieffer-Heeger model. We then discuss the notion of complex energy gaps and how they lead to intrinsically non-Hermitian phenomena. Next, we present the non-Hermitian skin effect, which is completely unique to non-Hermitian systems. We subsequently highlight the broken bulk-boundary correspondence in non-Hermitian topological phases and how the role of a generalized Brillouin zone becomes crucial, using the non-Hermitian Su-Schrieffer-Heeger model as a concrete example. The role of disorder and its interplay with symmetries in non-Hermitian systems is next summarized. In particular, we highlight the intriguing phenomenon of non-Hermitian Anderson skin effect. The linear response theory is widely used for characterizing the response of any system under the influence of an external perturbation. In this regard, a framework of non-Hermitian linear response has been recently proposed, which we present. We further analyze the transport signatures in the non-Hermitian Chern insulator as a concrete example of the theory. We then present the notion of non-equilibrium steady states in non-Hermitian systems, which entails a significant departure from the Hermitian case. We next survey the numerous experimental advances made in the field. Finally, we end with an outlook on the possible upcoming new directions in this rapidly developing field.

\begin{figure*}
\includegraphics[scale=0.30]{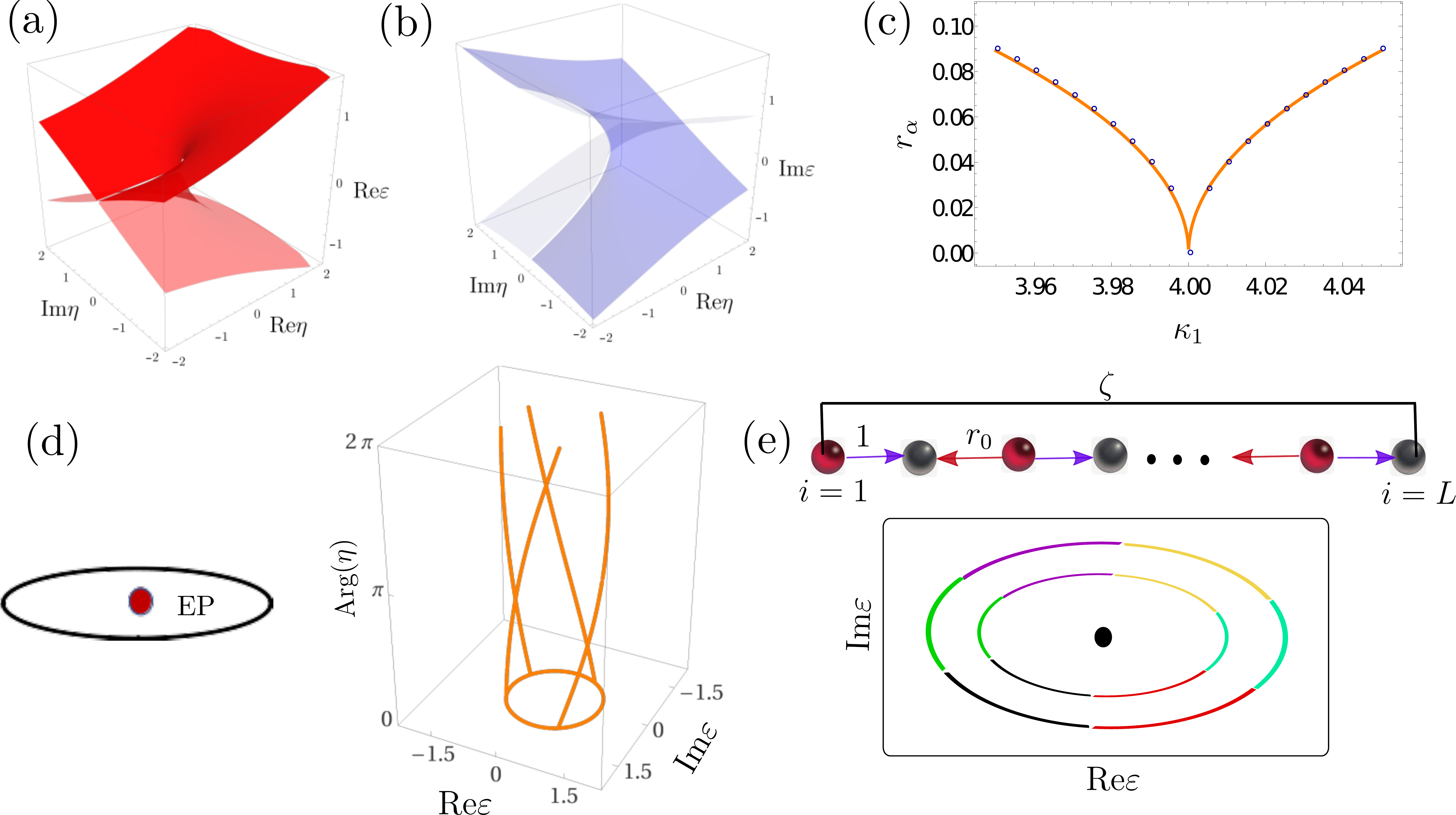}
  \caption{\textbf{Characterization and topological features of exceptional points.} Perspective view of (a) real and (b) imaginary energy exhibiting Riemann sheet structure of two coalescing energy levels in the complex $\eta$-plane. (c) Phase rigidity for diagnosing EPs in the parameter space showing square root scaling around the EP. (d) Eigenvalue switching manifests the Riemann sheet topology while encircling a fourth-order EP by tuning the system parameters. (e) Top panel is a schematic describing the model in Eq.~\ref{ep-bbc}. The bottom panel is a schematic of eigenvalue spectra in different colors showing how the eigenvalues vary with $\phi$ in the complex plane for different values of $\zeta$. The six colors in the spectrum represent three unit cells. As long as $\zeta \neq 0$, one needs three rounds of $\phi$ to return to the initial eigenvalue sheet. Here $\zeta = 0$ is a higher-order EP, where all eigenstates coalesce.} \label{ep-figure}
\end{figure*}

\section{Exceptional Points} \label{eps}

One of the most striking feature of non-Hermitian systems is the  presence of non-Hermitian degeneracies called exceptional points (EPs)~\cite{kato2013perturbation}. These EPs are spectral singularities in the parameter space (which is in general complex) of the Hamiltonian where both the eigenvalues and the eigenvectors coalesce. This is unlike Hermitian degenaracies where the eigenmodes become orthogonal at a Dirac point~\cite{hasan2010colloquium}. We start with a simple two-level system that reveals the intricate structures of EPs illustratively. Consider the Hamiltonian matrix,

\begin{equation}
       H  = \begin{pmatrix}
        \epsilon_1 & \kappa_1\\
        \kappa_2  & \epsilon_2
\end{pmatrix}.\label{ep-matrix}
\end{equation}

Note that this Hamiltonian describing this two-level system is non-Hermitian $(H^{\dagger}\neq H)$ since $\kappa_1 \neq \kappa^{*}_2 $, in general. The energy eigenvalues can be written as $\varepsilon_{\pm}=E_0 \pm \sqrt{\eta}$, where $E_0=\dfrac{ \epsilon_1+ \epsilon_2}{2}$ and $\eta=\dfrac{(\epsilon_1- \epsilon_2)^2}{4}+ \kappa_1 \kappa_2$. In general all parameters $\epsilon_1,\epsilon_2,\kappa_1$ and $\kappa_2$ can be complex. Interestingly, $\eta=0$ is a degeneracy point. We will see how this goes beyond a usual Hermitian degeneracy (see Fig.~\ref{ep-figure}(a)). At this special point, we investigate the nature of eigenmodes. We will invoke the notion of bi-orthogonal basis consisting of both left and right eigenvectors~\cite{brody2013biorthogonal}. Let $|\psi^R\rangle$ and $\langle \psi^L|$ be the right and left eigenvectors of $H$ with the same eigenvalue $\varepsilon_i$ such that as long as $\varepsilon_i \neq \varepsilon_j$, they become bi-orthogonal $\langle \psi_{i}^L|\psi_{j}^R\rangle=\delta_{ij}$. We have
 
 \begin{equation}
     H |\psi^R\rangle = \varepsilon_i |\psi^R\rangle,  \quad     \langle \psi^L| H = \varepsilon_i \langle \psi^L|.
 \end{equation}
 
The right and left eigenvectors of the system read
 
\begin{equation}
    |\psi^R\rangle = \begin{pmatrix}
\dfrac{(\epsilon_1- \epsilon_2) }{2\kappa_2} \pm \dfrac{\sqrt{\eta}}{\kappa_2} \\
1
\end{pmatrix}, \quad  \langle \psi^L| = \begin{pmatrix}
\dfrac{(\epsilon_1- \epsilon_2) }{2\kappa_1} \pm \dfrac{\sqrt{\eta}}{\kappa_1} & 1\\
 \end{pmatrix}.
\end{equation}

Strikingly, at the degeneracy point $\eta=0$ both the right eigenvectors and left eigenvectors coalesce. Consequently, the eigenvectors do not span the full Hilbert space and the system becomes defective -- this is the key signature of an EP. The right and left eigenvectors are orthogonal to each other at the EP. There exists a quantitative measure of eigenfunctions’ biorthogonality. It is the phase rigidity, which is an experimentally measurable quantity. The phase rigidity, $r_{\alpha}$, is defined as~\cite{rotter2009non,rotter2001dynamics},

\begin{equation}
    r_{\alpha} =\dfrac{\langle \psi_{\alpha}^L|\psi_{\alpha}^R\rangle}{\langle \psi_{\alpha}^R|\psi_{\alpha}^R\rangle},
\end{equation}
 where $\psi^L_{\alpha}$ and $\psi^R_{\alpha}$ are the  biorthogonal left and right eigenvectors of a state $\alpha$. On approaching the EP $(\eta \rightarrow 0)$, right and left eigenvectors align and the phase rigidity vanishes $(r_{\alpha}\rightarrow 0)$ (see Fig.~\ref{ep-figure}(c)). In contrast, $r_{\alpha} = 1$ for a Hermitian system. In summary, EPs are a special kind of non-Hermitian degeneracy in the complex parameter space of the Hamiltonian where two or more eigenstates coalesce and the spectrum collapses to a single eigenvalue showing spectral singularities. Mathematically, the algebraic multiplicity (degenerate eigenvalues) exceeds the geometric multiplicity (number of independent eigenvectors) with an incomplete set of eigenfunctions, rendering the Hamiltonian defective, a situation that is truly unique to non-Hermitian system~\cite{heiss2012physics}. 
 In general, a Hamiltonian with dimension $m$ being non-diagonalizable at the EP with energy $\varepsilon_0$, can be brought to a Jordan block form of order $m$. The system still has $m$ degenerate eigenvalues, but there exists only one eigenvector, indicating an $m$-th order EP ~\cite{heiss2008chirality}. It is to be noted that, in general, an $m \times m$ matrix (representing an $m$-dimensional Hamiltonian) can have an EP of any order $n \leq m $. This corresponds to a block diagonal form, where one of the blocks is comprised of an $n \times n$ Jordan block, and the other block is a $(m-n) \times (m-n)$ diagonal matrix.

The crucial distinction between the Hermitian and exceptional degeneracies can be effectively captured by rewriting our Hamiltonian in the following generic form~\cite{bergholtz2021exceptional}

\begin{equation}
H(\delta) = \mathbf{f} (\delta)\cdot \sigma,
\label{two band model}
\end{equation}

 where $\mathbf{f}= \mathbf{f}_R + i \mathbf{f}_I $ with $f_R, f_I \in \mathbb{R}^3$ and the Pauli matrices form the basis of two dimensional matrices and represent the pseudo spin degrees of freedom. Hermitian degeneracies occur with the condition that $\mathbf{f}_R=0$, which can be tuned by the parameters (three real constraints) of spatial dimension in a three-dimensional system which will typically be a point. In general, the co-dimension of the band crossing (degeneracies) in $m$ dimensions would be $p = m -m_{BC}$, where $m_{BC}$ is the dimension of the band crossing (for instance, $m_{BC} = 0$ for point crossings and $m_{BC}= 1$ for line crossings) ~\cite{chiu2016classification}. We note that Hermitian degeneracies in non-Hermitian systems require simultaneous satisfaction of $\mathbf{f_I}=0$ and $\mathbf{f_R}=0$, thus leading to the tuning of six real parameters. Whereas, in Hermitian systems, $\mathbf{f_I}=0$ trivially.
  For the present case, the Hermitian degeneracy occurs only if both $\epsilon_1=\epsilon_2$ and $\kappa_1= \kappa^{*}_2=0$ -- the resulting degeneracy has co-dimension 3. Whereas, for the non-Hermitian case $\mathbf{f}_I$ is in general non-zero. The occurrence of non-Hermitian degeneracies leads to the following conditions --

\begin{equation}
f_R^2-f_I^2 =0 \quad \text{and} \quad \mathbf{f}_R \cdot \mathbf{f}_I =0,
\label{two band constraint}
\end{equation}

with the tuning of two spatial dimensions. One may also encounters trivial degeneracies, also known as diabolic points, with both $\mathbf{f}_R = \mathbf{f}_I =0$. These lack the information regarding coalescence of eigenfunctions~\cite{shen2018topological}.

It is interesting to note that exceptional contours (ECs) can also appear in a non-Hermitian system. These are comprised entirely of EPs, where phase rigidity vanishes over a surface instead of a single point~\cite{cerjan2018effects}. The general criteria for obtaining an EC are the following

\begin{equation} \label{eq1}
\begin{split}
 \text{Re det}[H (\delta)] & = 0, \\
  \text{Im det}[H (\delta)] & =0.
\end{split}
\end{equation}

These two constraint equations in a higher dimensional system allow to form EC where both the eigenvectors and eigenvalues merge without any additional symmetry. These ECs preserve quantized topological charges~\cite{parto2021non,cerjan2016exceptional}. Please note that the above two conditions spontaneously lead to the vanishing phase rigidity criterion, which is essential for an exceptional manifold ~\cite{cerjan2018effects}. We will discuss about the tuning of these exceptional contours through non-hermitian strength and periodic driving in section \ref{semimetals} and \ref{floquet-eng} respectively.

Very recently, the role of higher-order EPs (HEPs) has come to the forefront in the context of enhanced response even with amplified perturbation~\cite{hodaei2017enhanced}. To better understand HEPs, let us consider a system described by a Hamiltonian $H_0$ subjected to a small perturbation $\lambda H_1$ in the vicinity of the EP. The eigenenergies of the combined system, $H(\lambda)=H_0+ \lambda H_1$, can be expressed as a power series in $\lambda$ in the neighborhood of the EP~\cite{kato2013perturbation,jaiswal2023characterizing}.

 \begin{equation}
 \varepsilon_h=\varepsilon+\alpha_1 \omega^h \lambda^{1/p}+\alpha_2 \omega^{2h} \lambda^{2/p}+...,
 \label{puisuex}
 \end{equation}
 
where $h=0,1,...,{p-1}$, $\omega=\exp(2\pi /p)$, $\alpha_{1}=\big(\langle\psi_L |H^{'}(0)|\psi_R\rangle\big)^{1/m}$, and $H^{'}(0)=\dfrac{\partial H(\lambda)}{\partial \lambda}|_{\lambda=0}$ and $p$ is the dimension of the system. It is worth noting that Eq.~\ref{puisuex} is a Puiseux series, well-known in mathematics. To provide readers with a comprehensive understanding of the Puiseux series and its geometric interpretation, we highly recommend referring to \cite{kato2013perturbation}, which offers an excellent introduction to this topic.

Such an expansion shows that $|\varepsilon_h(\lambda)-\varepsilon|$ is in general of the order of $|\lambda|^{1/p}$ for small $\lambda$. This immediately suggests that the first-order response of a non-Hermitian system varies as $|\lambda|^{1/p}$ around an EP. Therefore, the system in a non-Hermitian setting seems to provide better sensitivity in sensors than the system in a Hermitian one, where the first-order response is expected to vary linearly in $\lambda$ ~\cite{parto2021non}. However, this issue remains somewhat controversial, and it has been reported in a few recent studies that the EP bears no dramatic sensitivity enhancement ~\cite{chen2019sensitivity,lau2018fundamental}. Since a $p$-th order EP requires coalescing of $p$ levels and results in a sudden reduction in dimensionality in the system, typically, there should be $2(p-1)$ real constraints needed to detect it in systems without any symmetry. However, recent pioneering studies reveal that symmetry can substantially reduce the number of constraints~\cite{mandal2021symmetry,sayyad2022realizing,delplace2021symmetry}. For a detailed discussion on symmetry classes and structure of EPs, we refer to section~\ref{semimetals} and~\ref{symmetry class}.

We note that Eq.~\ref{puisuex} reveals delicate topological behaviour around an EP. For instance, the fractional exponent in Eq.~\ref{puisuex} suggests that the different eigenmodes intersect each other due to the underlying Riemann sheet topology when enclosing an EP by tuning the system parameters. Specifically, if we encircle an EP (without intersecting it) with a closed loop $S$, then the eigenvalues can be expressed as a set of $p$ holomorphic functions such as $\{ \varepsilon_1(\lambda), \varepsilon_2(\lambda), ..., \varepsilon_p(\lambda) \}$ (see Fig.~\ref{ep-figure}(d)). Now if $S$ is rotated continuously around $\lambda = 0$, these $p$ functions can be continued analytically~\cite{heiss2016circling,kato2013perturbation,jaiswal2023characterizing}. They would have undergone a permutation among themselves upon one revolution of $S$. For instance, in our case of Eq.~\ref{ep-matrix}, we consider a small perturbation away from the EP $(\eta=0)$ by a complex number $(\eta=\Delta e^{i \phi}$, where $\Delta,\phi \in \mathbb{R})$ in the parameter space. When we tune $\phi$ from $0$ to $2\pi$ to encircle the EP in the parameter space, the two eigenvalues get exchanged around the EP -- $\{E_0+\sqrt{\eta},E_0-\sqrt{\eta}\}\rightarrow \{E_0-\sqrt{\eta},E_0+\sqrt{\eta}\} $ (see Fig.~\ref{ep-figure}(a)). This exchange behavior has been geometrically interpreted in terms of the holonomy matrix~\cite{ryu2012analysis}. The intricate topological structure in the neighbourhood of multiple EPs leads to interesting consequences~\cite{gao2018nature}. The topological nature stems from the fact that these rich non-trivial scenarios do not depend on the precise shape of the loop as long as it encloses the EP.

We next delve into the emergence of the topological properties arising from the appearance of EPs while transitioning from periodic boundary conditions (PBC) to open boundary conditions (OBC) through the tuning of a twisted boundary condition. To analyze this we adapt the following model from Ref.~\cite{xiong2018does} by choosing $\epsilon_1=\epsilon_2=0$, $\kappa_1=1$ and $\kappa_2=r_0 e^{-i k}$, such that Eq.~\ref{ep-matrix} becomes

\begin{equation}
    H  = \begin{pmatrix}
        0 & 1\\
       r_0 e^{-i k}  & 0
\end{pmatrix}.
\label{ep-bbc}
\end{equation}

This model gives rise to an EP at $r_0=0$, and the swapping of eigenvalues can be observed around the EP by tuning $k$ from $0$ to $2\pi$. Next, we go to the real space version of this model after a Fourier transformation, where $r_0$ describes the unidirectional inter-unit cell hopping and unity on the right corner in Eq.~\ref{ep-bbc} represents the unidirectional intra-cell hopping in the unit cell comprising of two inequivalent atoms, as depicted in Fig.~\ref{ep-figure}(d). Furthermore, we modulate the boundary condition with a twisted coupling between the first and the last sites. The system Hamiltonian in real space representation takes the following form

\begin{align}
H =  & \sum_{i=1}^L c^\dagger_{i,A} c_{i,B}  +
r_0 c^\dagger_{i+1,A} c_{i,B} + \zeta e^{i \phi} c^\dagger_{L,B} c_{1,A},  
\label{Ham-bbc-real}
\end{align}

where $c^{\dagger}_{i,\alpha} (c_{i,\alpha})$ is the fermionic creation (annihilation) operator at site $i$ for sublattice $\alpha=A,B$ (see Fig.~\ref{ep-figure}(d)). Here $L$ is the total number of sites. The last term specifies the twisted hopping between two ends. $\zeta=0$ corresponds to OBC. Whereas with $\zeta=r_0$ and $\phi=0$ the translation symmetry is restored and the eigenvalues read $\varepsilon= r_0 e^{i k /2}$ with $L$ discrete $k$ points. Here, $\phi$ connects the wavevectors through a gauge transformation -- it transforms all the related phases between inter unit cells to a total phase jump $e^{i \phi} = e^{i L k}$ across the two ends~\cite{niu1985quantized}. One can tune $\phi$ from $0$ to $2\pi$ in Eq.~\ref{Ham-bbc-real} to observe eigenmodes switching among $L$ eigenvalues in the complex plane (see Fig.~\ref{ep-figure}(d)). Now, if we modulate the boundary condition with $\zeta=\Theta  r_0$ and unless $\Theta$ approaches to zero the translational symmetry can still be maintained with a renormalized hopping. Consequently, the swapping of eigenvalues in a complex loop can still be observed (see Fig.~\ref{ep-figure}(d)). Strikingly, at $\Theta=0$ we are right at an EP under OBC and all the eigenstates coalesce to a \emph{single} point in the spectrum. Consequently, the eigenspectrum changes dramatically while tuning from PBC to OBC. Thus, the connection between bulk topological invariant in PBC and the presence of topological boundary modes becomes elusive and leads to violation of celebrated bulk boundary correspondence~\cite{bergholtz2021exceptional,xiong2018does}, which we discuss in detail in Sec. \ref{generalized Brillouin Zone}. The interplay between the role of higher-order EPs, extreme sensitivity in boundary conditions, and the occurrence of bulk-boundary correspondence is a very active topic of current research~\cite{kunst2019non,zhang2022review,martinez2018topological}.

\section{Paradigmatic Models} \label{models}

Next, we discuss key models which have been instrumental in the development of topological ideas in the context of non-Hermitian systems -- the Hatano-Nelson model, the non-Hermitian Su-Schrieffer-Heeger model, the non-Hermitian Chern insulator model, and non-Hermitian topological semimetal models. In the following we present the interesting spectral topology and topological band properties of each of these systems, which serve as the foundation for the following discussions.

\subsection{Hatano-Nelson Model}
  
\begin{figure*}
\includegraphics[scale=0.44]{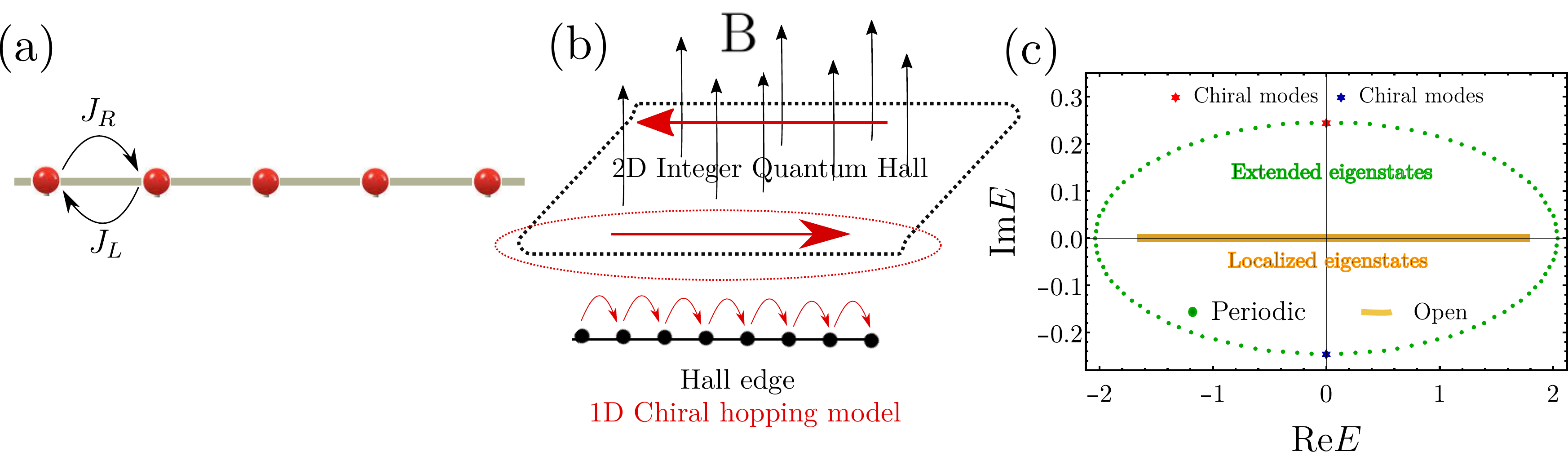}
  \caption{\textbf{Hatano-Nelson model.} Schematic illustration of the Hatano-Nelson model (see Hamitonian in Eq.~\ref{hatano-hamiltonian}). (b) The one-dimensional system characterized by an integer winding number in the chiral hopping model corresponds to the boundary of a two-dimensional system characterized by an integer quantum Hall state with a quantized Chern number. (c) The dispersion of the model in the complex plane under periodic and open boundary conditions. The chiral modes are also indicated, along with extended and localized eigenstates.} \label{Hatano-figure}
\end{figure*}

We start with the Hatano-Nelson model, a prototypical one band non-Hermitian model with non-reciprocal hopping in a one-dimensional geometry, which was originally proposed to study localization and vortex pinning in disordered type II superconductors~\cite{hatano1996localization,hatano1997vortex}. The non-interacting Hatano-Nelson model has the Hamiltonian

\begin{equation}
  H=\sum_j (J_L c_j^{\dagger}c_{j+1}+J_R c_{j+1}^{\dagger}c_{j}),
  \label{hatano-hamiltonian}
\end{equation}
    
where $J_R, J_L \in \mathbb{R}$ introduce the hopping asymmetry in the lattice leading to the non-Hermiticity (see Fig.~\ref{Hatano-figure}(a)). Moving to the $k$-space by Fourier transforming the operators (using $c_j= \sum_k c_k e^{i k j}$), we obtain the momentum space Hamiltonian, $H_k=(J_R + J_L) \cos{k}+ i (J_R - J_L) \sin{k}$. The dispersion is an ellipse in the complex plane with symmetric real energy and asymmetric imaginary energy parts, stipulating the non-reciprocal nature of the system. The eigenbands parameterized by $k$ wind around the origin in anticlockwise (clockwise) for positive (negative) $J_L-J_R$. The different winding in the two cases can be described by spectral winding number $\omega$,

\begin{equation}
 \omega  = \int^{\pi}_{-\pi} \dfrac{d k}{2 \pi i} \partial_k \ln (E_k-E_B),
 \label{winding-formula}
\end{equation}
  
which is nothing but the number of times the complex eigenband encloses the base point with energy $E_B$. The winding number reads

\begin{equation}
\begin{split}
\omega & = 1 \quad \textrm{for} \quad  |J_L| > |J_R|, \\
       & = -1 \quad \textrm{for} \quad |J_L| < |J_R|.
\end{split}
\end{equation}

Here $J_L= J_R$ is a topological phase transition point. This structure of winding number suggests that the system exhibits a point gap around the origin $(E_B=0)$, which we will discuss in more detail. We note that in a remarkable recent study it has been shown that the Hatano-Nelson model corresponds to the edge of a two-dimensional system characterized by an integer quantum Hall state with a Chern number under appropriate time dynamics (see Fig.~\ref{Hatano-figure}(b))~\cite{lee2019topological,franca2022non}. Interestingly, this insight has led to the understanding that the winding numbers have a close correspondence with the chiral modes. The single particle current can be determined by the group velocity $\left(\propto \dfrac{\partial \text{Re}[E_k]}{\partial k}\right)$ and a lifetime given by the inverse of the imaginary part of the energy eigenvalue~\cite{lee2019topological}. The left (right) propagating wave contributing to the chiral current depends on the sign of the positive (negative) imaginary energy. In general, the one-to-one correspondence between winding number and chiral modes can be found as follows
 
 \begin{equation}
\omega =\dfrac{1}{2} \sum_{n\alpha} \text{sign}\big[\text{Im}\big(E_n(k_{n \alpha})\big)]\text{sign}\big[\partial_k[\text{Re}\big(E_n(k_{n \alpha})\big)\big],
 \end{equation}
 
where $\{k_{n \alpha} : \text{Re}\big(E_n(k_{n \alpha})\big)=0\}$, i.e., $k_{n \alpha}$ are the wavevectors set where the real part of the energy vanishes, and $\alpha$ labels the number of such connected points in the spectrum. Here $n$ denotes the band index. Thus, the topological winding number of the complex spectrum counts the number of chiral modes with $\text{Re}\big(E_n(k_{n \alpha})\big)=0$. For instance, in case of the Hatano-Nelson model, the winding number corresponds to the difference between left-propagating modes minus the number of right-propagating modes for $k=\pi/2$. We will further discuss the notion of nonequilibrium phases arising from single-particle spectrum of the model in section~\ref{many-body}.
 
Strikingly, another unique feature of this non-Hermitian model is the extreme sensitivity to the boundary conditions. The spectrum having completely real dispersion under OBC drastically changes under PBC where we find a complex spectrum in general (see Fig.~\ref{Hatano-figure}(c)). Interestingly, the nature of eigenstates also changes radically since, under PBC, we get extended eigenstates. In contrast, the eigenstates become localized with an exponential amplitude profile in OBC, leading to the non-Hermitian skin effect. Consequently, under OBC, a macroscopic number of eigenstates accumulate at one of the edges with a dominant directional hopping dependence, a manifestation of non-reciprocal hopping. We will discuss this aspect in more detail subsequently. Moreover, at the strong non-reciprocity limit $(J_R$ or $J_L \rightarrow 0)$, higher-order EPs appear with an algebraic multiplicity scaling with system size while the geometric multiplicity becomes unity. This indicates that all the bulk modes align to one state, which is exponentially localized at the edge under OBCs, leading to the violation of the celebrated bulk boundary correspondence. We will come back to these aspects later in this review (see section~\ref{generalized Brillouin Zone}).
  
\subsection{Non-Hermitian Su-Schrieffer-Heeger Model}

Next, we discuss another important non-Hermitian model, namely the non-Hermitian Su-Schrieffer-Heeger (SSH) model. It is a one-dimensional tight-binding model on a bipartite lattice with non-reciprocal intra-unit cell hopping and $PT$ symmetric imaginary staggered potential~\cite{wu2021topology,lieu2018topological,herviou2019defining,yao2018edge}. We note that several other variants and generalizations also exist in the literature~\cite{xie2019topological,perez2018ssh,fu2020extended,zhang2021topological,li2014topological,schomerus2013topologically,rudner2009topological,kartik2022multi}. Notably, the parent Hermitian SSH model is paradigmatic in its own right with several intriguing topological properties~\cite{su1979solitons,asboth2016short}. The non-Hermitian SSH model Hamiltonian on a finite chain with $L$ sites reads $H=H^{\text{hop}}+H^{\text{pot}}$, where

\begin{align}
H^{\text{hop}} = - & \sum_{i} [t_1 (c^\dagger_{i,A} c_{i,B} + h.c.) +
t_2 (c^\dagger_{i+1,A} c_{i,B} + h.c.)+ t_3 (c^\dagger_{i+1,B} c_{i,A} + h.c.)]  \notag \\ +& \sum_{i}
\delta (c^\dagger_{i,B}c_{i,A} - c^\dagger_{i,A}c_{i,B}),
\label{eqn:Ham-ssh-1}
\end{align}

and 
\begin{align}
H^{\text{pot}} = i \sum_{i}
\gamma (c^\dagger_{i,A}c_{i,A} - c^\dagger_{i,B}c_{i,B}).
\label{eqn:Ham-ssh-2}
\end{align}

\begin{figure*}
\includegraphics[scale=0.39]{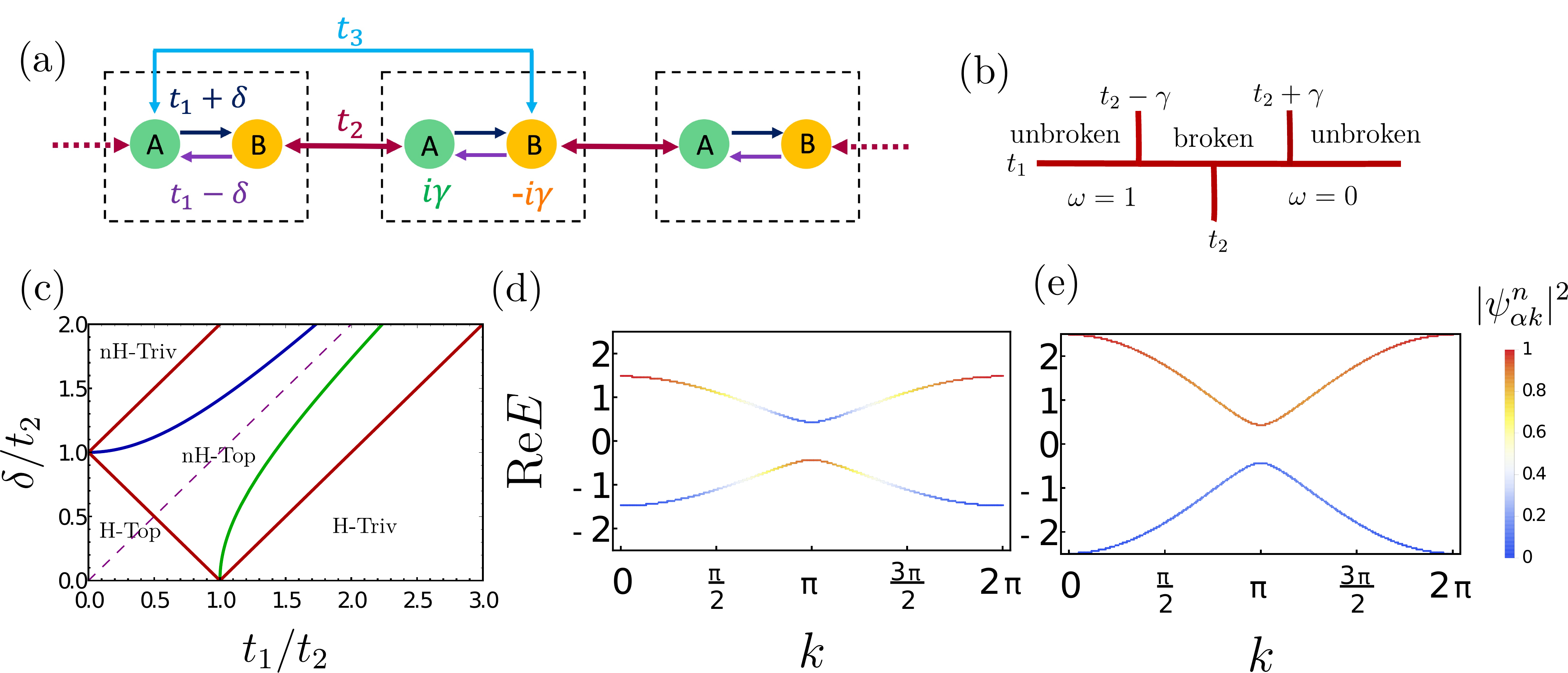}
  \caption{\textbf{Non-Hermitian Su-Schrieffer-Heeger model.} (a) Schematic depiction of the generalized non-Hermitian Su-Schrieffer-Heeger model. Sublattice atoms $A$ and $B$ form the unit cell. The intra unit cell hopping is non-reciprocal, given by $t+\delta$ and $t_1-\delta$ from left to right and right to left, respectively. The inter unit cell hopping strength is $t_2$. Next-nearest neighbour hopping potential from sublattice $A$ to $B$ is given by $t_3$. A non-Hermitian gain and loss has also been included on sublattices $A$ and $B$, with strength $\gamma$. Here, $\delta$ and $\gamma$ are the non-Hermitian parameters of the model.  (b) The one-dimensional phase diagram with $t_1$ for the $PT$-symmetric case consists of $PT$ broken and unbroken phases as a function of $t_1$. Here $\omega$ is the topological index that counts the pairs of gapless-real-energy edge modes in the system. (c) Periodic (red) and open boundary condition (blue, green, and purple-dashed) phase diagram of the model in the absence of staggered potential $(\gamma=0)$ derived from the single-particle gap closings. (d) and (e) The distinction between H-top and H-triv phases through orbital band character. The real part of the dispersion is shown with $|\psi|^2$ contribution in color with $t_1 = 0.5$ and $\delta = 0.25$ (d), and $t_1 = 1.5$ and $\delta = 0.25$ (e). Note the difference in the orbital character around $k = \pi$ for the two cases. We choose $t_2=1.0$, $t_3=0$, and $\gamma=0$.} \label{SSH-figure}
\end{figure*}

Here $c^{\dagger}_{i,\alpha} (c_{i,\alpha})$ is the fermionic raising (lowering) operator at site $i$ for sublattice $\alpha=A,B$ (see Fig.~\ref{SSH-figure} (a)). Here $t_1$ and $t_2$ denote the intra- and inter-unit cell hopping amplitudes, respectively, and $\delta$ introduces a non-reciprocity only in the intra-unit cell hopping, thereby introducing non-Hermiticity in the system. Here $t_3$ represents the next-nearest neighbor hopping from sublattice $A$ to $B$. We first consider $t_3 = 0$ for simplicity. The onsite imaginary potential with balanced gain-and-loss $\pm i  \gamma$ incorporates $PT$ symmetry in the model. We present a detailed discussion of different symmetry classes in different parameter regimes of the model in the next section (see section~\ref{symmetry class}).

Since the system possesses translational symmetry under PBC, the eigenvectors and eigenvalues can be parameterized by $k$. In $k$ space, the Hamiltonian can be written as

\begin{equation} 
H(k) =
\begin{pmatrix}
i\gamma & t_2 e^{-i k}+t_1-\delta\\
t_2 e^{i k}+t_1+\delta & -i\gamma 
\end{pmatrix}.
\label{specexpan}
\end{equation}    
    
In the Hermitian limit, $\gamma=\delta=0$. For both $t_1 < t_2$ and $t_1 > t_2$, the dispersion is gapped with a bond dimer-like order and different polarizations, which can be characterized by a Bloch topological invariant or winding number~\cite{asboth2016short}. Here $t_1=t_2$ is a gapless critical point separating these two gapped phases with different orbital characteristics. The topological phase with a nontrivial winding number gives rise to zero energy edge modes under OBC. Next, we consider two distinct cases of the non-Hermitian model parameters and discuss the topological features of both regimes.

(a) Symmetric intra-unit cell hopping $(\delta=0)$ in the presence of onsite potential $(\gamma \neq 0)$: In this case, the model exhibits both $PT$ symmetry broken and unbroken phases. Interestingly, the $PT$ symmetry induces unique topological phase transition in the system. Unlike the Hermitian SSH model, the edge modes can exist in the entire $PT$ symmetry broken region as long as $t_1< t_2$ even with bulk band crossing. The phase transition happens when bulk band merges with edge albeit with an imaginary gap in the dispersion.The topological phase transition is governed by a topological index, which a complex Berry phase can characterize~\cite{lieu2018topological}. We present the complete phase diagram in this phase in Fig.~\ref{SSH-figure}(b). \\

(b) Non-reciprocal hopping with non-zero $\delta$ in the absence of staggered potential $(\gamma=0)$: In this case, the Hamiltonian respects the sub-lattice symmetry $S: c_{iA} \rightarrow -c^\dagger_{iA}, c_{iB} \rightarrow c^\dagger_{iB}$, $S  H^\dagger  S^{-1} = H $, and $\sigma_zH(k)\sigma_z^{-1}=-H(k)$~\cite{kawabata2019classification}. Consequently, the energy eigenvalues come in $\pm$ pairs at each $k$. We follow the notations introduced in Ref.~\cite{herviou2019defining} and discuss the detailed features of the system. The PBC phase diagram is divided into four regions considering the single-particle spectrum gap closings (for absolute values of energies) for lines $\delta= t_1 \pm 1$ and $\delta=1-t_1$ (see Fig.~\ref{SSH-figure}(c)): (i) Hermitian topological $(\delta < (1-t_1))$ -- the complex spectrum has a two-lobe structure with a real energy line gap in the complex plane $\{\text{Re}[E], \text{Im}[E]\}$. In subsequent Section~\ref{complex energy gap}, we present a detailed discussion of various kinds of complex energy gaps. (ii) Hermitian trivial $(t_1 > 1+ \delta)$ -- spectral topology shows the structure similar to (i) (see Fig.~\ref{Fig: Energy}(a)). Interestingly, both phases (i) and (ii) show distinctive features in terms of orbital band character, revealing their topological and trivial nature, respectively~\cite{banerjee2022chiral} (see Fig.~\ref{SSH-figure} (d) and (e)). (iii) Non-Hermitian topological $(\delta > (|1-t_1|), \delta< (1+t_1))$ -- the spectral topology changes to a single loop in the complex plane from a two-lobe structure in (i) via Lifshitz transitions (see Fig.~\ref{Fig: Energy} (b)). (iv) Non-Hermitian trivial $(\delta>(1+t_1))$ -- the complex spectrum comprises of two spectral lobes vertically displaced along the imaginary axis with an imaginary line gap. The phase transitions between different phases are marked by EPs with a significant change in spectral topology. The spectral topology of four regions [(i)-(iv)] are characterized by winding numbers (see Eq. \ref{winding-formula}) $(1, 1, 1/2, 0)$, respectively. Interestingly, similar to the Hatano-Nelson model, the existence of chiral modes and the finite winding numbers are interconnected and guaranteed by the symmetry-protected spectral topology~\cite{kawabata2019symmetry,lee2019topological,yoshida2021correlation}. For OBC, one finds four distinct regions as shown in Fig.~\ref{SSH-figure} (c). Briefly, $\delta>t_1$ region leads to real eigenvalues, while for $\delta < t_1$, the eigenvalues are complex. Moreover, within the region $ \sqrt{t^2_1-1} < \delta < \sqrt{t^2_1+1}$, one finds topological phases characterized by zero energy edge modes which are perturbatively connected to topological boundary modes in the Hermitian limit ($\delta=0, t_1<t_2$). The immediate incongruity between the PBC and OBC phase diagram stems from the non-Hermitian nature of the system, reflecting the extreme sensitivity to the boundary conditions. We will subsequently present various interesting features of the model, considering the intricate interplay of symmetry, spectral topology, and boundary sensitivity~\cite{lee2020many,herviou2019defining,jin2019bulk,kunst2018biorthogonal,liu2019topological,esaki2011edge,schomerus2013topologically,alvarez2018non,yin2018geometrical,st2017lasing,borgnia2020non,vyas2021topological,he2020non,silberstein2020berry} (see section~\ref{generalized Brillouin Zone}  and~\ref{many-body}).

\subsection{Non-Hermitian Chern Insulators}

We now introduce a two-dimensional non-Hermitian Chern insulator model, which has been important in understanding higher dimensional non-Hermitian phases. We will explain its topological characterization in terms of the non-Hermitian generalization of the Chern number following pioneering work in Refs.~\cite{yao2018non,shen2018topological}. We will also discuss the transport properties of the model in section~\ref{transport}. The momentum space Hamiltonian of the system reads

\begin{equation}
    H (\mathbf{k}) = (v_x \sin{k_x} + i \gamma_x) \sigma_x + (v_y \sin{k_y} + i \gamma_y) \sigma_y + (m - t_x \cos{k_x} - t_y \cos{k_y} +i \gamma_z) \sigma_z.
    \label{chern insulator}
\end{equation}

The Hermitian part of the model corresponds to the Qi-Wu-Zhang model~\cite{qi2006topological}, whereas the non-Hermitian parameters $\gamma_i$ $(i=x,y,z)$ describe the imaginary coupling between the orbitals, which can be thought as a manifestation of imaginary Zeeman field~\cite{lee1952statistical}. In the Hermitian limit, $\gamma_{x,y,z}=0$, the model exhibits topological phase transitions at $m=t_x+t_y$, where the Chern number switches. 

Before we go to the topological characterization of this Chern insulator model, we take a brief detour and discuss the topological band theory for a non-Hermitian Hamiltonian with the generalized notion of gapped band structures in the complex parameter plane introduced by Shen \emph{et al.}~\cite{lee1952statistical}. For a non-Hermitian Hamiltonian of a periodic system, the Bloch theorem suggests that the energy eigenvalues can be parameterized by crystal momentum $\mathbf{k}$ in the Brillouin zone, thus defining a band structure. A band labelled by index $n$ is called "gapped" or "separable" if its energies in the complex plane do not overlap with that of any band. More precisely, one can define a band to be separable if its energy $E_n(\mathbf{k}) \neq E_m(\mathbf{k}) $ for all $m \neq n$ and all $\mathbf{k}$. In contrast, a band is called inseparable if at some momentum $\mathbf{k}$ the complex energy $E_n(\mathbf{k})$ overlaps with another band $E_m(\mathbf{k})$ and becomes degenerate. We can now define the non-Hermitian version of the Chern number defined over the Brillouin zone for an energy band in two dimensions, which serves as the basis of topological classification in non-Hermitian systems. For a separable band structure $(\langle \psi_{n}^L|\psi_{n}^R\rangle \neq 0)$, with energy $E_n$, the Chern number can be defined as~\cite{shen2018topological}

\begin{equation}
C_n^{\Theta \nu} = \dfrac{1}{2 \pi} \int_{\mathrm{BZ}} \epsilon_{i j} B^{\Theta \nu}_{n, ij} (\mathbf{k}) d^2 \mathbf{k},
\end{equation}

where $\epsilon_{ij}=-\epsilon_{ji}$ and $ B^{\Theta \nu}_{n, ij} (\mathbf{k})=i\langle \partial_i  \psi_{n}^{\Theta}(\mathbf{k})|\partial_j  \psi_{n}^{\nu}(\mathbf{k})\rangle$, with the normalization condition $\langle \psi_{n}^{\Theta} (\mathbf{k})|\psi_{n}^{\nu} (\mathbf{k})\rangle =1$. Here $\Theta,  \nu= L/R$, where $L/R$ represent the left and right eigenvectors. Interestingly, Ref.~\cite{shen2018topological} shows that all four combinations of Chern numbers are equal $C^{LL}=C^{LR}=C^{RL}=C^{RR}$, characterizing the same topological features of a band.

\begin{figure*}
\includegraphics[scale=0.43]{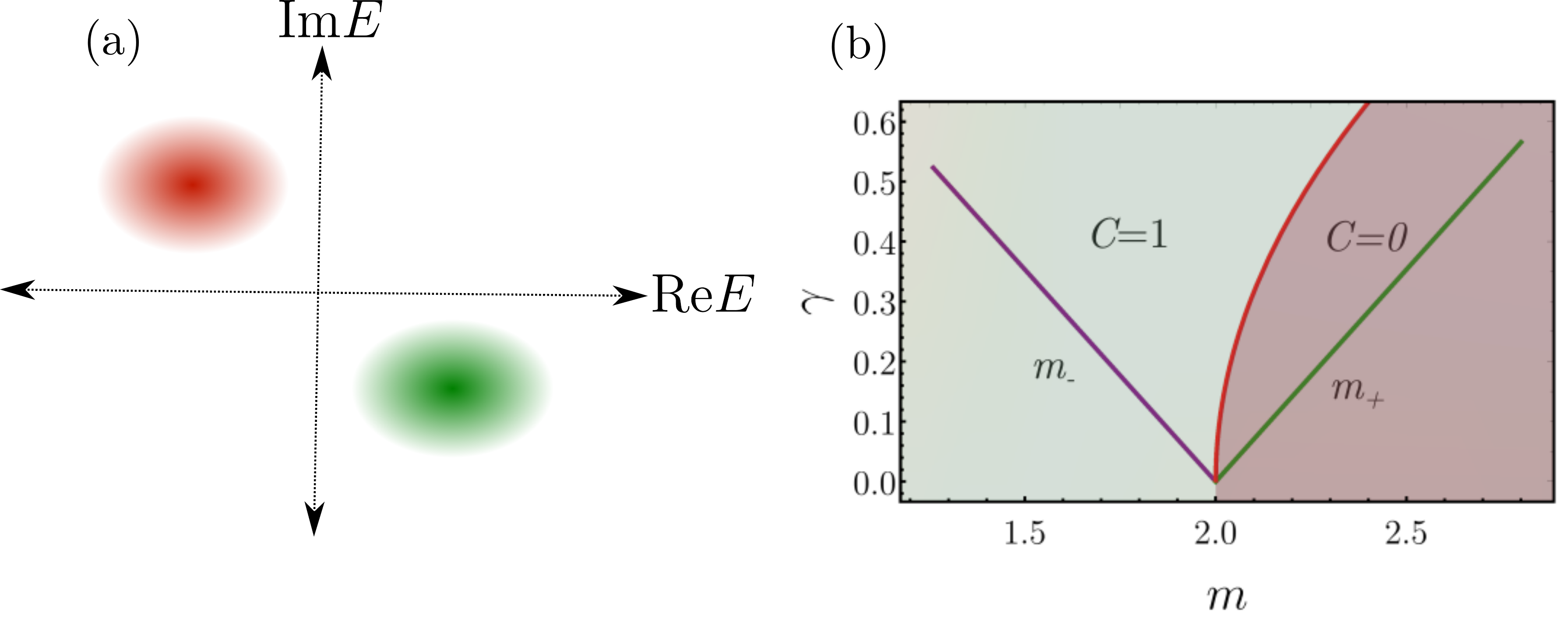}
  \caption{\textbf{Non-Hermitian Chern insulator.} (a) The notion of separable bands. The energies of the two bulk bands are shown in red and green regions. (b) Topological phase diagram based on open boundary spectra, Bloch, and non-Bloch Chern numbers. The Bloch-Hamiltonian phase boundaries are shown as the green and solid purple lines denoted by straight lines $m = m_{\pm}$. The non-Bloch Chern number $C$, defined in Eq.~\ref{non-bloch-chern}, predicts the presence of non-trivial chiral edge states in the blue-shaded area with the non-Bloch Chern number $C=1$. The phase boundary dividing the topological and trivial regions can be well approximated by the red line denoted by the curve $m=2+\gamma^2$.} \label{Chern-figure}
\end{figure*}

Coming back to the non-Hermitian Chern insulator model, we find that the Bloch spectrum is gapped for $m>m_{+}$ and $m<m_{-}$ and the Chern number becomes $0$ and $1$ respectively. Here $m_{\pm}=t_x+t_y \pm \sqrt{\gamma^2_x+\gamma^2_y}$ with $\gamma_z=0$. The gap closes at the Bloch phase boundaries for $m=m_{\pm}$ and the Chern number is undefined in between these phase boundaries (see Fig.~\ref{Chern-figure} (b)). Strikingly, the Bloch Chern number dramatically fails to mimic the edge state information of the corresponding open boundary system. The dissimilarity between open boundary and periodic boundary Bloch spectra results in the violation of conventional bulk-boundary correspondence~\cite{xiong2018does}. The resolution to this puzzle comes from the non-Bloch band theory proposed by Yao \emph{et al.}~\cite{yao2018non}, which will be discussed in detail in Section~\ref{generalized Brillouin Zone}. Briefly, the non-orthogonal nature of eigenvectors suggests an exponential profile of bulk eigenstates, which in turn demands a modification in the Bloch wavevector with an imaginary component. The formalism goes as follows. The low energy continuum model for the Chern insulator up to order $k_j^2$ can be written as

\begin{equation}
\begin{split}
    H(\mathbf{k}) & = (v_x k_x + i \gamma_x) \sigma_x + (v_y k_y + i \gamma_y) \sigma_y+ (m-t_x-t_y+\dfrac{t_x}{2}k_x^2+\dfrac{t_y}{2}k_y^2)\sigma_z,\\
    & = H_0+H_1,
    \end{split}
\end{equation}

where $H_1=i \gamma_x \sigma_x+ i \gamma_y \sigma_y$ and $H_0$ is the remaining part of the Hamiltonian. For small $\mathbf{k}$, $H_1$  can be thought as the first-order response of $H_0$, i.e, $H_1=i \sum_{j=x,y}\dfrac{\gamma_j}{v_j}\dfrac{\partial H_0}{\partial k_j}=\sum_j \dfrac{\gamma_j}{v_j}[x_j,H_0]$. This problem can be solved considering $H_1$ as a perturbation. The lowest order perturbation to an eigenfunction of $H_0$ takes the form $\exp[(\gamma_x /v_x)x +(\gamma_y /v_y)y]$. This suggests that we need a complex-valued wavevector to describe the open boundary eigenstate

\begin{equation}
    \mathbf{k} \rightarrow \mathbf{\Tilde{k}} + i \mathbf{\Tilde{k'}},
\end{equation}

where $\Tilde{k_j}=-\gamma_j/v_j$ for small $\mathbf{\Tilde{k'}}$ in this model. One can write the non-Bloch Hamitonian as

\begin{equation}
\Tilde{H}(\mathbf{\Tilde{k}})=H(\mathbf{k}\rightarrow\mathbf{\Tilde{k}} + i \mathbf{\Tilde{k'}}).
\end{equation}

Now one can define the non-Bloch Chern number in terms of left and right eigenvectors parameterized by $\mathbf{\Tilde{k}}$   

\begin{equation}
\Tilde{C_n^{\Theta \nu}} = \dfrac{1}{2 \pi} \int_{\Tilde{T^2}} \epsilon_{i j} \Tilde{B^{\Theta \nu}_{n, ij}}(\mathbf{\Tilde{k}}) d^2\mathbf{\Tilde{k}},
\label{non-bloch-chern}
\end{equation}

over the generalized Brillouin zone $\Tilde{T^2}(\mathbf{\Tilde{k}})$. The non-Bloch Chern number can precisely predict the topological boundary states under OBC with topological phase boundary at $m=2+\gamma^2$ separating the two distinct phases with non-Bloch Chern numbers 0 and 1 (see Fig.~\ref{Chern-figure} (b)). 

\subsection{Non-Hermitian Topological Semimetals}  \label{semimetals}  
 
In recent years the concepts of topology have been extended to gapless systems, such as nodal semimetals~\cite{nielsen1983adler,armitage2018weyl,murakami2007phase,burkov2011weyl,burkov2011topological,young2012dirac}. They host topologically protected stable degeneracies, which lead to intriguing properties owing to their distinctive electronic excitations. The effect of non-Hermiticity in such systems has been an emerging area of research in the last years. The presence of EPs in the dispersion fundamentally changes the nature of nodal semimetals resulting in various kinds of exceptional manifolds, such as lines, rings, surfaces, and complex nexus structures in different dimensions~\cite{xu2017weyl,yoshida2019symmetry,cerjan2019experimental,ghorashi2021non,ghorashi2021non,liu2021higher,rui2019pt,zhang2019experimental,zhou2019exceptional,tang2020exceptional,wang2021simulating,he2020double,liu2021higher}. In particular, the role of unitary and antiunitary symmetries has been studied in the context of the stability and intricate structure of the higher-order exceptional manifolds in various dimensions ~\cite{staalhammar2021classification,delplace2021symmetry,mandal2021symmetry,sayyad2022realizing}.  Remarkably, the generalized notion of the energy gap in a complex plane introduces multiple topological structures with distinct gapless phases truly unique to non-Hermitian systems~\cite{kawabata2019classification,zhen2015spawning,molina2018surface,moors2019disorder,okugawa2019topological}. For instance, due to topological protection of EPs, one can obtain an open Fermi arc around an EP with fractional topological charge in direct contrast to conventional Dirac and Weyl semimetals~\cite{zhou2018observation,bergholtz2019non}. Interestingly, the ramified symmetry classification in non-Hermitian systems enriches the catalog of symmetry protected topological phases~\cite{kawabata2019symmetry,budich2019symmetry,shiozaki2021symmetry,sayyad2022symmetry}. Recent studies show that various non-Hermitian spatial symmetries, which act non-locally in the momentum space, can stabilize an exceptional unconventional Weyl semimetal by enforcing general constraints on band degeneracies~\cite{rui2022non}. Ref.~\cite{budich2019symmetry} has put forward important examples of symmetry-protected topological phases governed by generic non-Hermitian symmetries, which act on the Hamiltonian in the following way $H= Q H^{\dagger} Q^{-1}, Q^{\dagger}Q^{-1}=QQ^{\dagger}=\mathbb{I}$. Interestingly, they find the increased dimensionality of the manifold of EPs (exceptional nodal surfaces) as compared to the case without symmetry. Additionally, we would like to note that a recent work (Ref.~\cite{staalhammar2021classification} ) has presented a comprehensive topological categorization of exceptional nodal degeneracies that are protected by $PT$ symmetry. This classification leads to the emergence of exceptional nodal topologies, such as second-order knotted surfaces with arbitrary genus, third-order knots, and fourth-order points.

\begin{figure*}
\includegraphics[scale=0.28]{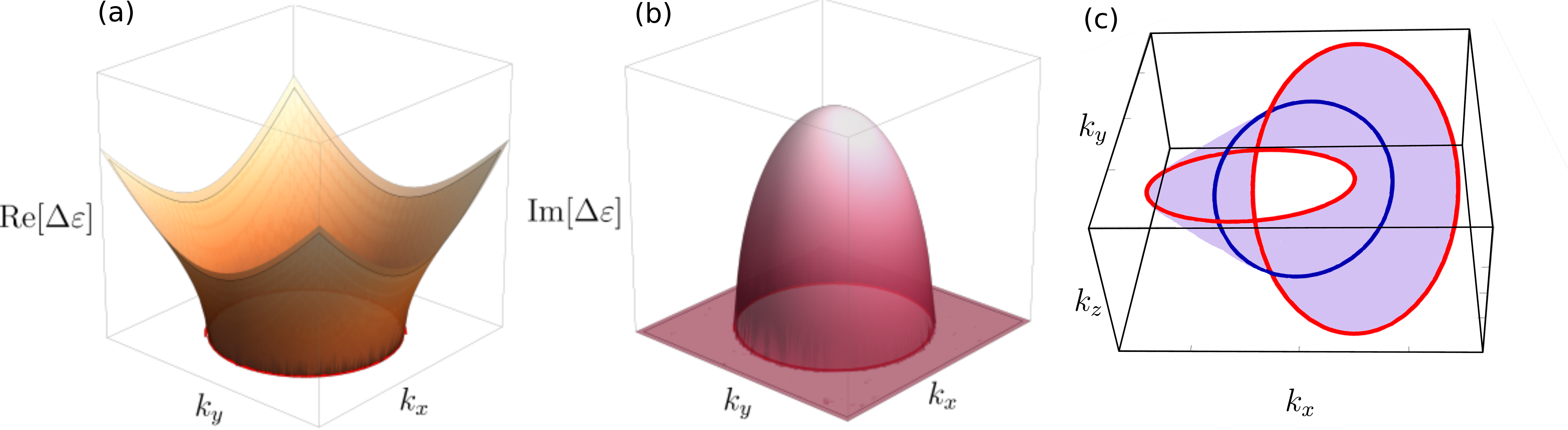}
  \caption{\textbf{Non-Hermitian topological metals.} (a) The real (orange) and (b) imaginary (magenta) parts of the spectral gap $\varepsilon$ as a function of momentum in a two-dimensional Weyl node in the presence of onsite gain and loss, illustrating a generic symmetry-protected non-Hermitian nodal phase. Contours of exceptional points are indicated by red lines, separating two disparate regions where the real energy is projected to zero (open Fermi volumes) from those where the imaginary part of $\varepsilon$ vanishes. The exceptional ring is symmetry protected with $\sigma_x H^{*} \sigma_x=H$, which constrains this particular kind of eigen structure of both components. (c) Illustration of a non-Hermitian Hopf-link exceptional line semimetal for the Hamiltonian of the form $H(\mathbf{k}) = [ \mathbf{h(k)} + i \lambda \mathbf{g (k)}] \cdot \mathbf{\sigma}$, with $h_x (\mathbf{k})=\sin{k_z},h_y (\mathbf{k})=0$ and $h_z (\mathbf{k})=m + \cos{k_x}+\cos{k_y}+\cos{k_z}$. The non-Hermitian part is chosen to be $g_x (\mathbf{k})=\sin{k_x}, g_y (\mathbf{k})=0$ and $g_z (\mathbf{k})=\sin{k_z}$. We set $m=-21/8$ and $\lambda=1/2$. The purple color represents the Fermi surface with a finite lifetime, whose boundaries (red curves) are Hopf-link exceptional lines. The blue line denotes the original nodal line on the purple Fermi surface. For more details, we refer to Ref.~\cite{yang2019non}.} \label{semimetal}
\end{figure*}

As we discussed previously, Eq.~\ref{two band model} and Eq.~\ref{two band constraint} suggest that tuning two parameters can lead to non-Hermitian phases with a substantial reduction in spatial dimension compared to their Hermitian counterpart. As an illustration, let us consider Eq.~\ref{two band model}, which respects $H= Q H^{\dagger} Q^{-1}$ with $Q=\sigma_x$. It trivially satisfies the condition $\mathbf{f}_R \cdot \mathbf{f}_I =0$, allowing a one parameter tuning to obtain an exceptional structure. Consequently, one obtains EPs, exceptional lines, and surfaces in one, two, and three-dimensional systems, respectively. The eigenvalues of the system read $\epsilon_{\pm}=\pm \sqrt{f^2_R-f^2_I}$ which suggests either purely real (or imaginary) eigen structures with non-Hermitian Fermi arcs manifesting the branch cut singularities connecting the EPs with a crucial distinction in dispersion~\cite{kozii2017non} (see Fig.~\ref{semimetal}). Consequently, non-Hermitian Fermi arcs, which are a bulk phenomenon, fundamentally differ from surface Fermi arcs in conventional Weyl semimetals~\cite{armitage2018weyl}. Interestingly, the two-dimensional Weyl node in the Hermitian phase with semi-metallic band structure turns into a metallic dispersion in the presence of EPs due to parameter redundancy~\cite{bergholtz2019non}. Very recently, topological semimetals have attracted much interest in terms of broken and un-broken bulk boundary correspondence under OBC~\cite{zhang2020bulk,jin2019bulk,kunst2018biorthogonal,edvardsson2019non,song2019non}. Among them, the topological features of non-Hermitian nodal semimetals~\cite{wang2019non}, semi-Dirac semimetals~\cite{banerjee2021non}, and Weyl semimetals~\cite{yang2022non} are worth mentioning.

Topological nodal line semimetals can display symmetry-protected knotted line degeneracies in momentum space in Hermitian systems~\cite{sun2017double,lian2016five,chen2017topological,yan2017nodal,chang2017weyl,ezawa2017topological,chang2017topological,zhou2018hopf,bi2017nodal}. Remarkably, non-Hermitian topological phases in three dimensions can give rise to knotted phases with robust band degeneracies even in the \emph{absence} of any symmetry~\cite{yang2019non,yang2020jones}. For instance, one can transit from nodal line semimetals to twisted Hopf link exceptional lines via Lifshitz transitions with distinctive changes in the Fermi surface characterized by a topological linking number without any symmetry constraint~\cite{yang2019non}. The knotted phase with metallic dispersion can be characterized by open Fermi surfaces with vanishing real energy gaps, known as Seifert surfaces, that are bounded by knotted lines of EPs~\cite{carlstrom2019knotted,carlstrom2018exceptional}. The eigenenergy strings of the knotted structures can provide complete information on topological classifications and concomitant phase transitions of non-Hermitian Hamiltonians with separable bands in terms of their respective knot invariants and braiding patterns~\cite{hu2021knots}. Interestingly, non-Hermitian nodal knot metals with delicate complex energy eigenbands lead to unique surface states, dubbed as \emph{tidal} surface states, which are related to band vorticity and layer structure of their dual Seifert surface, revealing the algebraic, geometric, and topological aspects of their parent knots~\cite{zhang2021tidal}.

Tuning of various topological semimetallic band structures has become an active topic of current research owing to their interesting geometry and fascinating transport properties both in Hermitian~\cite{kitagawa2011transport,narayan2015floquet,yan2016tunable,narayan2016tunable,toudert2017interband,hubener2017creating,xu2019tuning,jaiswal2020floquet,bao2022light,li2018realistic} and non-Hermitian systems~\cite{wu2022non,he2019floquet,zhou2021floquet,zhou2022q,zhou2022driving,banerjee2022emergent,li2020topological,chowdhury2022exceptional}. Hermitian Hamiltonians hosting Weyl points with arbitrary topological charge can be transformed to one-dimensional exceptional contours in the presence of non-Hermitian gain-and-loss perturbations~\cite{cerjan2018effects,chowdhury2022exceptional} (please see the discussion on exceptional contours in section ~\ref{eps}). Tuning of such non-Hermitian perturbations leads to a new class of topological phase transitions via merging of oppositely charged exceptional contours~\cite{cerjan2018effects}. For example, in Ref.~\cite{zyuzin2018flat}, the authors discuss the disorder-driven topological phase transitions in type-II Weyl semimetals, where type II Weyl points transform into a flat band or a nodal line segment depending on the tilt direction. Recent studies show that symmetry-protected two-dimensional non-Hermitian topological phases can be realized by tuning the staggered asymmetric hopping strengths in one-dimensional superlattices~\cite{hou2021two}. On the other hand, non-Hermitian systems subjected to periodic driving can exhibit topological phases that are not possible in their static counterparts. One approach to stimulating non-Hermitian topological phases through periodic driving is to use Floquet engineering, which we will discuss in the next section.

\section{Floquet Engineering of non-Hermitian topological phases} \label{floquet-eng}

 Floquet engineering -- tuning properties of systems using periodic driving -- is a useful tool to manipulate quantum matter~\cite{oka2019floquet,oka2009photovoltaic}. In the standard picture, the time-evolution operator of a Floquet system is periodic and can be decomposed into a series of unitary operators, known as Floquet operators, which correspond to the time-evolution of the system over one period of the driving. In particular, Floquet topological phases are a fascinating class of topological matter that arise in periodically-driven, or Floquet systems. These time-periodic systems offer unique topological features since the eigenspectra winding of the Floquet operator, known as quasienergy, has the same periodicity as the time-periodic driving ~\cite{oka2009photovoltaic}. We point the reader to Refs.~\cite{rudner2019floquet,bukov2015universal} for an excellent introduction to Floquet engineering and its topological features.

Recently, there has been a growing interest in combining the ideas of Floquet engineering with the notion of non-Hermitian topological phases. Non-Hermitian Floquet topological phases have received significant attention due to their unique properties and potential applications in various fields. Strikingly, time-periodic non-Hermitian systems feature complex band energies, invalidating the Wannier-Stark localization~\cite{oka2019floquet}, a fundamental feature of Floquet topological phases. This requires a restructuring of the framework of non-Hermitian time-periodic systems which similarly restores the Wannier-Stark localization using non-unitary Floquet theory, as non-Bloch band theory does with the GBZ~\cite{wang2022non-floquet}.

We briefly sketch the formalism of Floquet theory for a non-Hermitian system. Consider a time-periodic system described by a non-Hermitian Hamiltonian $H(t) = H(t + T)$ with period $T$ in a complete set of basis $|u_{\alpha}(t)\rangle$. Next, let us define a non-unitary time evolution operator, i.e., Floquet operator $U(t_1,t_2)$ which governs the time translation from time $t_1$ and $t_2$. Then, the Floquet theorem dictates

\begin{equation}
    U(t_1+n T,t_0)= U(t_1,t_0)[U(t_0+T,t_0)]^n.
\end{equation}

For the stroboscopic analysis $(t_0=0)$, the Floquet operator can be written as

\begin{equation}
    U(T,0)=\sum_{\alpha} e^{-i \varepsilon_{\alpha} T} |u_{\alpha}(0)\rangle \langle u_{\alpha}(0)|,
\end{equation}

where $|u_{\alpha}(0)\rangle$ and $\varepsilon_{\alpha}$ are called Floquet states and complex quasienergies, respectively. Then the time-periodic system can be determined by the Floquet equation

\begin{equation}
    [H(t) -
i \partial_t]|\psi_{\alpha}(t)\rangle = \varepsilon_{\alpha}|\psi_{\alpha}(t)\rangle,
\label{floquet}
\end{equation} 

such that the Floquet modes evolve as

\begin{equation}
|\psi_{\alpha}(t)\rangle = \sum_{\alpha} c_{\alpha} e^{-i \varepsilon_{\alpha} t} |u_{\alpha}(0)\rangle.
\end{equation}

The Floquet equation (Eq.~\ref{floquet}) defines an effective non-unitary Floquet Hamiltonian $H_{\text{eff}} =\dfrac{i}{T} \ln{U(T,0)}$ with eigenvalues $\varepsilon_{\alpha}$. One can expand the Floquet modes in a Fourier series in multiplies of the driving frequency $(\omega=2\pi/T)$,

\begin{equation}
 |\psi_{\alpha}(t)\rangle =   e^{-i \varepsilon_{\alpha} t} \sum^{\infty}_{m=-\infty} e^{-i m \omega t} \phi_{\alpha,m}.
\end{equation}

\begin{figure*}
\includegraphics[scale=0.55]{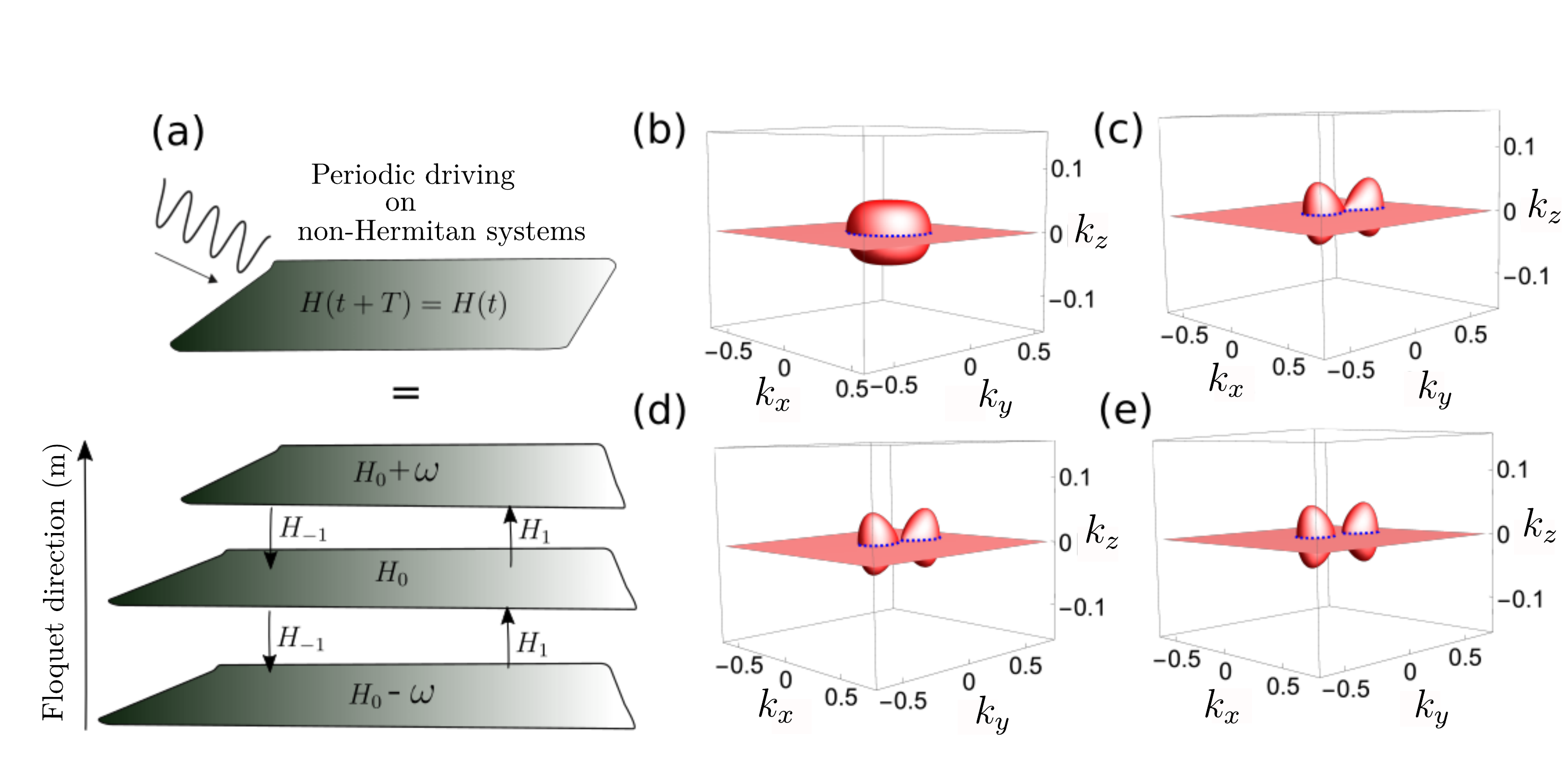}
  \caption{\textbf{Floquet engineering of non-Hermitian topological phases.} (a) Schematic representation of Floquet Hamiltonian. Floquet theory transforms the time-periodic system into a time-independent layered one-body problem in which Floquet mode indices $(m)$ represent different layers. The inter-layer hopping is specified by $H_0$ and the black arrow $H_m$ $(m\neq 0)$ denotes the temporal hopping. Further, $m\omega$ introduces a fictitious static electric field in the Floquet direction. Under the high-frequency approximation, the different layers become decoupled, and the system exhibits Wannier-Stark localization. (b)-(e) Floquet engineering of non-Hermitian Weyl semimetal leads to Lifsitz transitions in the exceptional contours with increasing light intensity. The system is described by the non-Hermitian Hamiltonian $H=\dfrac{1}{2m}\big(k^2_{-} \sigma_{+}+k^2_{+} \sigma_{-}\big)+ (\eta v_z k_z + i \zeta)\sigma_z$ in the presence of a circularly polarized field of the form $A(t) = A_0 (\cos{\omega t},0,-\sin{\omega t})$. The single exceptional contour (the blue dotted line) gradually splits into two, resulting in a Lifshitz transition under Floquet driving. We set $\zeta=0.05, k_z=0.0$ and $\omega=v_z=\eta=1.0$. Here the amplitude of driving changes as $A_0=0.0, A_0=0.445,A_0=0.448$ and $A_0=0.51$, from (b)-(e). We refer to Ref.~\cite{chowdhury2021light} for more details.} \label{floquet-2}
\end{figure*}

Next, we substitute this in the Floquet equation (Eq. \ref{floquet}) to obtain a time-independent eigenvalue equation of the form

\begin{equation}
\sum^{\infty}_{m=-\infty}    H_p  
 \phi_{\alpha, m-p} = (\varepsilon_{\alpha}+ m \omega) \phi_{\alpha,m},
\end{equation}

where $H=\sum^{\infty}_{p=-\infty} e^{-i p \omega t} H_{p}$. The index $\alpha$ labels the eigenstates, while $m$ and $p$ are the Fourier mode indices. Consequently, in single-particle problems, this boils down to an effective model with one extra dimension. The index $p$ of the Fourier mode represents the lattice site index with intra- and inter-layer couplings in a fictitious (temporal) “Floquet direction” (see Fig.~\ref{floquet-2} (a)). It can be shown that under a high-frequency approximation, the different layers become decoupled, and the state shows the Wannier-Stark localization along the Floquet direction. Furthermore, one can write down an effective time-independent Hamiltonian as follows

\begin{equation}
    H_{\text{eff}}(\textbf{k})=H_0+ \sum_{m \geq 1} \dfrac{[H_{+ m},H_{- m}]}{m \omega} + \mathrm{}{O()},
\end{equation}

where $H_{0}$ is the bare Hamiltonian and $H_{±m} =\dfrac{\omega}{2 \pi} \int^{T}_{0}
H(t)e^{\pm i m\omega t}dt $ are Fourier components of the time-dependent Hamiltonian. The underlying Floquet theory has been successfully incorporated in characterizing as well as predicting unique non-Hermitian topological phases~\cite{zhou2018non,koutserimpas2018nonreciprocal,zhou2021dual,zhou2019dynamical,hockendorf2020topological,zhou2020non,hockendorf2019non,wu2021floquet-second,liu2022symmetry}. For instance, the Floquet driving scheme redefines the notion of EPs leading to ``Floquet Exceptional Points (FEPs)", which correspond to the merging of two or more quasi-energies and corresponding Floquet eigenmodes of a time-periodic non-Hermitian Hamiltonian~\cite{longhi2017floquet,zhang2020non}. Furthermore, the interplay of FEPs and non-Hermitian dynamics has also gained attention~\cite{longhi2017floquet}. In Ref.~\cite{wu2020floquet}, the authors proposed a generalised scheme to recover the bulk boundary correspondence in the light of non-Hermitian topological phases under periodic driving even with broken chiral symmetry and disorder. The study of the Floquet non-Hermitian skin effect (FNHSE) and intriguing aspects of anomalous edge modes has  become a current theme of active research in the context of boundary sensitivity of non-Hermitian systems~\cite{wu2020floquet,wu2021floquet}. Furthermore, in Ref.~\cite{li2019topological} the authors show that non-Hermitian Floquet topological insulators can host non-reciprocal dissipationless edge modes as well as regimes of decaying and amplifying topological edge transport. In an interesting recent work by Zhou \emph{et al.}~\cite{zhou2022q}, the authors proposed a framework for constructing an interesting class of topological matter, which exhibit non-Hermiticity induced fractional quasi-energy corner modes and NHSE with fractional quasi-energy edge states in Floquet open systems. The study of light-matter topological insulators with non-Hermitian topology also dubbed as ``Floquet exceptional topological insulator" in various dimensions has been emerging as an active topic of current research in the context of non-Hermitian skin effect, and its concomitant point gap topology as well as associated wave dynamics~\cite{dash2023floquet}.

It has been recently shown that periodic driving may be used to control the position and stability of EPs in non-Hermitian topological semimetals~\cite{banerjee2020controlling}. One can simulate non-Hermitian semimetals exhibiting exceptional rings through driving. Associated topological phase transitions arising from the application of light can be traced out by the nontrivial Hall conductivity~\cite{he2020floquet}. Furthermore, recent studies show that driving can be used as a facile tool to generate exceptional contours (see Fig.~\ref{floquet-2} (b)-(e)) (we discussed about exceptional contours in section ~\ref{eps}). The charge division of the created exceptional contours and the concomitant Lifshitz transitions have been studied in non-Hermitian multi-Weyl semimetals in the presence of circularly polarized light~\cite{chowdhury2021light}. One can also tailor and control other higher-order topological phases exhibiting two-dimensional corner and three-dimensional hinge states by tuning the gain and loss or the periodic driving~\cite{wu2021floquet,luo2019higher,ghosh2022non}. Recently, the study of non-Hermitian quasi-crystals has become a point of theoretical and experimental interest, given the immense body of work that has been dedicated to this particular topic~\cite{liu2021localization,longhi2019topological,liu2020non,chen2022quantum,acharya2022localization,dai2022dynamical}. The Floquet driving protocol enriches the topological features in non-Hermitian quasi-crystals revealing unique localization transitions. The topological winding number based on Floquet quasi-energies has been employed to characterize the non-Hermitian quasi-crystalline phases with different localization nature~\cite{zhou2021floquet,zhou2022driving}. Another interesting study of driven dissipative quasicrystal systems was the discovery of topological triple-phase transitions featuring the intertwining of topology, symmetry breaking and mobility phase transitions in coupled optical fibre loops~\cite{weidemann2022topological}. Floquet engineering of non-Hermitian topological matter thus seems to be a promising future avenue in synthesizing unique non-equilibrium phases in open systems~\cite{zhou2021floquet,zhou2022driving,zhang2020non,banerjee2022emergent}.

\section{Non-Hermitian Symmetry Classes}\label{symmetry class}

The well-known Altland-Zirnbauer classification of Hermitian topological insulators and superconductors~\cite{altland1997nonstandard}, which segregates Hermitian systems into 10 distinct symmetry classes based on the presence or absence of three internal symmetries -- time-reversal symmetry (TRS), particle-hole symmetry (PHS) and chiral symmetry (CS) -- is modified in the presence of non-Hermiticity~\cite{lee2019topological,luo2022unifying,shiozaki2021symmetry,liu2019topological,hamazaki2020universality,gong2018topological,vecsei2021symmetry,liu2022symmetry}, first pointed out by Bernard and LeClair \cite{bernard2002classification}. Non-Hermiticity results in the ramification and, in some cases, a unification of these symmetry classes, as shown by Kawabata \textit{et al.}~\cite{kawabata2019symmetry}. The fundamental difference which leads to recasting the symmetry classification is that for non-Hermitian systems $H^\dagger \neq H$ and hence, $H^T \neq H^{*}$. This leads to the bifurcation of PHS and CS under non-Hermiticity. To illustrate, a Hermitian system with PHS follows $C H^* C^{-1}=-H$, where $C$ is a unitary operator. Since complex conjugation and transposition are not equivalent operations for non-Hermitian systems, PHS gets ramified as $C H^* C^{-1}=-H$ and $C H^T C^{-1}=-H$. Similarly, for Hermitian systems CS follows $\Gamma H \Gamma ^{-1} = -H$; under non-Hermiticity, this separates into $\Gamma H \Gamma ^{-1} = -H$ and $\Gamma H^\dagger \Gamma ^{-1} = -H$ where $\Gamma$ is a unitary operator. Moreover, TRS being an antiunitary symmetry follows either of the following in Hermitian systems, $T_+ H^* T_+^{-1} = H$ or $T_- H^* T_-^{-1} = -H$. However, a non-Hermitian Hamiltonian $H$ can be directly mapped to $iH$ with a one-to-one correspondence. Hence, $T_+ H^* T_+^{-1} = H $ implies $T_+ (iH)^* T_+^{-1} = -(iH)$, making $T_+ \equiv T_-$, unifying both symmetries. 
These internal symmetries of a non-Hermitian system ($C$, $T$ and $\Gamma$ such that $CC^*=\pm 1$, $TT^*=\pm 1$, and $\Gamma^2=1$), along with pseudo anti-Hermiticity (PAH) -- $\eta H^\dagger(k)=\eta^{-1} = -H(k)$ where $\eta^2 = 1$ -- form the generators of the non-Hermitian symmetry classification~\cite{lieu2019non,zhou2019periodic,gong2018topological,budich2019symmetry,lieu2018topological}. The interplay of non-Hermiticity and these discrete symmetries leads to a 38-fold symmetry classification summarized in Table~\ref{TableSymm}. Further, more recent studies show that time-independent non-Hermitian systems with a line gap, as well as time-dependent Floquet non-Hermitian systems can be classified into 54 different non-Hermitian symmetry classes, as shown in Refs.~\cite{liu2019topologicaldef,liu2022symmetry,ashida2020non}. 

One of the interesting consequences of this modification of symmetry classes is that despite topological phases being impossible in the one-dimensional class AI for Hermitian systems, the presence of non-Hermiticity leads to a possibility of topologically protected edge states. Remarkably, these zero-mode energy states in the non-Hermitian one-dimensional symmetry class AI have been experimentally observed in photonic lattices~\cite{weimann2017topologically}.
The classification of non-Hermitian symmetries also leads to the systematic prediction of symmetry-protected topological lasers, as shown in Ref.~\cite{kawabata2019symmetry}, which can have several physical applications.

As a concrete illustration, we present the symmetries in the generalized non-Hermitian SSH model described previously in Fig.~\ref{SSH-figure}(a). We tabulate each symmetry along with their constraint equations and symmetry operators in Table~\ref{TableNHSSH} and highlight the conditions under which it is possible to obtain the specific symmetry. By imposing certain constraints on the system parameters, it is possible to transition between different symmetry conditions.\\
Different symmetries combined with the nature of complex energy gaps in the system spectra lead to various interesting topological phenomena. We discuss more about these in the following sections.

\renewcommand{\tabcolsep}{0.17cm}
\begin{table}
\centering
\caption{\label{TableSymm} \textbf{Classification of non-Hermitian symmetries.} Combinations of each group of symmetries (i.e. each column) give rise to 10 symmetry classes giving a total of $4\times 10= 40$ symmetry classes. However, there are some redundancies in counting. For example, the symmetry-less class A has been counted 4 times -- once in each group. The correct combinations accounting for redundancies gives rise to the 38-fold classification of non-Hermitian symmetries. Here, TRS$_c$ and TRS$_t$ denote the symmetry considering conjugation and transposition of $H$ respectively. Similarly, for PHS$_c$ and PHS$_t$.}

\begin{tabular}{cccc}
 \hline
 \hline

  TRS$_c$ : $H= T H^* T^{\dagger}$ & TRS$_t$ : $H= T H^T T^{\dagger}$ & TRS$_c$ : $H= T H^* T^{\dagger}$ & TRS$_t$ : $H= T H^T T^{\dagger}$ \\

 PHS$_c$ : $H= -C H^* C^{\dagger}$ & PHS$_t$ : $H= -C H^T C^{\dagger}$ & PHS$_t$ : $H= -C H^T C^{\dagger}$ &  PHS$_c$ : $H= -C H^* C^{\dagger}$\\

 CS: $H=-\Gamma  H \Gamma^{-1}$ & CS: $H=-\Gamma  H \Gamma^{-1}$ & PAH : $H=-\eta H^\dagger \eta^{-1}$ & PAH : $H=-\eta H^\dagger \eta^{-1}$ \\

  \hline
  \hline
\end{tabular}
\end{table}

\renewcommand{\tabcolsep}{0.17cm}
\begin{table}
\centering
\caption{\label{TableNHSSH} \textbf{Symmetry conditions for generalized non-Hermitian Su-Schrieffer-Heeger model.} Each symmetry has its own constraint equation which is satisfied by the symmetry operator acting on the Hamiltonian. The conditions on the system parameters required to preserve the symmetry is given in the last column. The symbol `$\times$' denotes that under no condition can the non-Hermitian SSH model preserve the corresponding symmetry.}

\begin{tabular}{cccc}
 \hline
 \hline
 
  Symmetry& Equation & Operator & Conditions \\
  \hline
  \hline

 PHS$_t$, $CC^*=1$ & $H(-k)= C H^T(k) C^\dagger$ & $\mathbb{I}_2$ & $t_2=t_3, \delta=0$\\
 
 PHS$_t$, $CC^*=-1$ & $H(-k)= C H^T(k) C^\dagger$ & $i \sigma_y$ & $\times$ \\
 
 TRS$_t$, $TT^*=1$ & $H(-k)= T H^T(k) T^\dagger$ & $\mathbb{I}_2$ & $t_2=t_3, \delta=0$\\
 
 TRS$_t$, $TT^*=-1$ & $H(-k)= C H^T(k) C^\dagger$ & $i \sigma_y$ & $\times$\\
 
 PHS$_c$, $CC^*=1$ & $H(-k)= -C H^*(k) C^\dagger$ & $\mathbb{I}_2$ & $\times$\\
 
 PHS$_c$, $CC^*=-1$ & $H(-k)= -C H^*(k) C^\dagger$ & $i \sigma_y$ & $t_2=t_3, \delta=0, \gamma=0$\\
 
 TRS$_c$, $TT^*=1$ & $H(-k)= T H^*(k) T^\dagger$ & $\mathbb{I}_2$ & $t_2=t_3, \gamma=0$\\
 
 TRS$_c$, $TT^*=-1$ & $H(-k)= T H^*(k) T^\dagger$ & $i \sigma_y$ & $\times$\\
 
 CS & $H(k)= -\Gamma H^\dagger(k) \Gamma^{-1}$ & $\sigma_z$ & $t_2=t_3, \delta=0$\\
 
 PAH & $H(k)= -\eta H^\dagger(k) \eta^{-1}$ & $\sigma_x$ & $\times$\\
 
 Pseudo-Hermiticity & $H(k)= \zeta H^\dagger(k) \zeta^{-1}$ & $\sigma_x$ & $t_2=t_3$\\
 
 Sublattice symmetry & $H(k)= -S H(k) S^{-1}$ & $\sigma_z$ & $\gamma=0$\\
 
 Parity & $H(-k)= P H(k) P^{-1}$ & $\sigma_x$ & $t_2=t_3, \delta=0,\gamma=0$\\
 
 Parity-time & $H(k)= (PT) H^*(k) (PT)^{-1}$ & $\sigma_x$ & $t_2=t_3, \delta=0$\\
 
 Parity-Particle hole & $H(k)= -(CP) H^*(k) (CP)^{-1}$ & $\sigma_x$ & $\times$\\
 
 Inversion & $H^\dagger(-k)= I H(k) I^{-1}$ & $\sigma_z$ & $\times$\\

  \hline
  \hline
\end{tabular}
\end{table}

\section{Complex Energy Gaps in non-Hermitian systems}\label{complex energy gap}

\begin{figure*}
    \centering
    \includegraphics[width=0.95\textwidth]{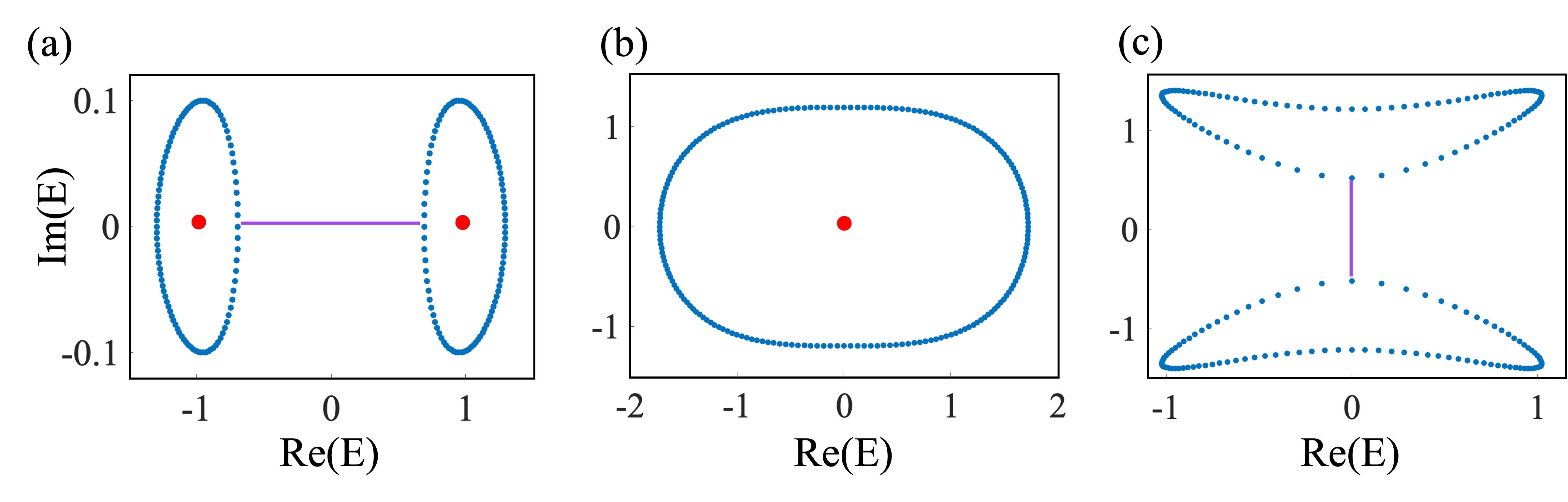}
    \caption{\label{Fig: Energy} \textbf{Point gaps and line gaps in the different phases in the non-Hermitian Su-Schrieffer-Heeger model.} (a) Prototype of the complex energy spectrum in the Hermitian phases (Hermitian topological or trivial phase) showing point gaps within the two spectral loops (shown by the red dots) and a real line gap separating energy loops (shown by the purple line). (b) The non-Hermitian topological phase has a point gap at zero energy shown by the red dot. (c) The non-Hermitian trivial phase which has an imaginary line gap between the two complex spectral loops (shown in purple).}
\end{figure*}

Unlike in Hermitian systems, the complex energy eigenvalues of non-Hermitian Hamiltonians cannot be ordered from lowest to highest, making the task of defining an energy gap non-trivial. Two kinds of complex energy gaps have been postulated to exist in non-Hermitian spectra~\cite{kawabata2019symmetry} -- point gap and line gap (illustrated in Fig.~\ref{Fig: Energy} for the non-Hermitian SSH model). A point gap can be defined as a point in the complex plane where no energy eigenstates exist. Unitary flattening of the spectrum cannot lead to crossing that particular point. Any non-Hermitian Hamiltonian with a point gap can be deformed to a unitary matrix which preserves the point gap and its symmetries.
On the other hand, a line gap is defined as an infinite line in the complex plane with no eigenenergy crossings. A line gap can, in general, be arbitrary. However, in the presence of TRS, the real axis forms the line gap, while under CS, the line gap becomes the imaginary axis. A non-Hermitian Hamiltonian with a line gap can be flattened to a Hermitian matrix which preserves the line gap and its symmetries. Hence, systems with a point-gap topology are inherently non-Hermitian with no correspondence to a Hermitian Hamiltonian. As such the presence of a point gap topology leads to intrinsically non-Hermitian phenomena~\cite{liu2019topological,borgnia2020non,zhong2021nontrivial,yoshida2021correlation,yoshida2022reduction}. We next discuss one such non-Hermitian phenomenon, namely the non-Hermitian skin effect.

\section{Non-Hermitian Skin Effect}

The non-Hermitian skin effect (NHSE) is a phenomenon completely unique to non-Hermitian systems, where a maximal number of bulk modes localize at the boundary of the system under OBC~\cite{yao2018edge,kunst2018biorthogonal,lee2019anatomy,zhang2022review,longhi2019probing,li2020critical,kawabata2020higher,yokomizo2021scaling,longhi2020unraveling,longhi2021non}. The NHSE has been theorised in several non-Hermitian systems in one~\cite{longhi2019probing, yokomizo2021scaling}, two~\cite{PhysRevB.104.125416,song2020two,sarkar2022non} and three dimensions~\cite{sun2021geometric,kawabata2020higher}. We note that the effect has also been recently observed experimentally in photonic systems~\cite{weidemann2020topological,song2020two,zhong2021nontrivial,zhu2020photonic,fang2022geometry}, acoustic topological insulators~\cite{zhang2021observation,puri2021tunable}, and topoelectrical circuits~\cite{zou2021observation,helbig2020generalized,yoshida2020mirror,hofmann2020reciprocal,galeano2022topological}, to highlight a few.

In order to intuitively understand the NHSE, we can consider the minimal non-Hermitian SSH model with non-reciprocal hopping, as shown in Fig.~\ref{SSH-figure}(a). We can ignore the on-site energies and the next nearest neighbour hopping $t_3$ for this discussion. The directionally-favoured hopping leads to a persistent current in the system under PBC, which results in the accumulation of the eigenstates at a boundary when OBC is imposed. This interrelation of the current in PBC and the NHSE in OBC was recently enunciated by Zhang \textit{et al.}~\cite{zhang2020correspondence}. The unit cell current under PBC can be calculated according to Ref.~\cite{asboth2016short} as

\begin{equation}
    J= \frac{iL}{2} [\langle c_j^\dagger c_{j+1} \rangle - \langle c_{j+1}^\dagger c_j \rangle], \label{eq:chiralcurrent}
\end{equation}

where $L$ is the length of the SSH chain, $c^\dagger_j$ and $c_j$ are the fermionic creation and annihilation operators at site $j$, respectively. Here $j$ denotes any lattice site in the chain since the same current flows throughout the system. For a one-dimensional system, the presence of a persistent current directly relates to a non-zero winding of eigenenergies in the complex plane as demonstrated in Ref.~\cite{zhang2020correspondence}, establishing that the non-zero winding number under PBC implies an NHSE under OBC and vice-versa. Further, the presence of either indicates a persistent current under PBC. The relationship between the skin effect with the winding number suggests that the NHSE originates due to the inherent topology of non-Hermitian systems, established by Okuma \textit{et al.}~\cite{okuma2020topological}. The necessary and sufficient conditions to obtain a skin effect under OBC can be summarised as follows: (a) in one-dimensional systems, the PBC complex energy spectrum must exhibit a point-gap topology~\cite{okuma2020topological}, and (b) in two or higher dimensions, the eigenspectrum under PBC must accommodate a finite spectral area in the complex plane~\cite{zhang2022universal}. 

Internal symmetries of the underlying non-Hermitian Hamiltonian can also play an interesting role in determining the nature of NHSE that the system demonstrates. For example, in systems preserving TRS, such as quantum spin Hall insulators, where the $\mathbb{Z}_2$ invariant becomes non-trivial, the eigenstates form Kramers pairs, one of which localizes at the left boundary while the other localizes at the right boundary under OBC. This type of symmetry-protected NHSE has been dubbed the $\mathbb{Z}_2$ skin effect~\cite{okuma2020topological}.
The interplay of non-Hermiticity and symmetries can also give rise to higher-order NHSE, where skin modes are localized at the corners of a sample rather than the edges. This has been exhibited in Refs.~\cite{luo2019higher,kawabata2020higher,zhang2021observation,okugawa2020second}. Furthermore, Lu \emph {et al.}~\cite{lu2021magnetic} studied the interplay of magnetic field and NHSE. Interestingly, they discovered that magnetic fields can strongly suppress the NHSE despite its topological origin.

\begin{figure*}
    \centering
    \includegraphics[width=0.8\textwidth]{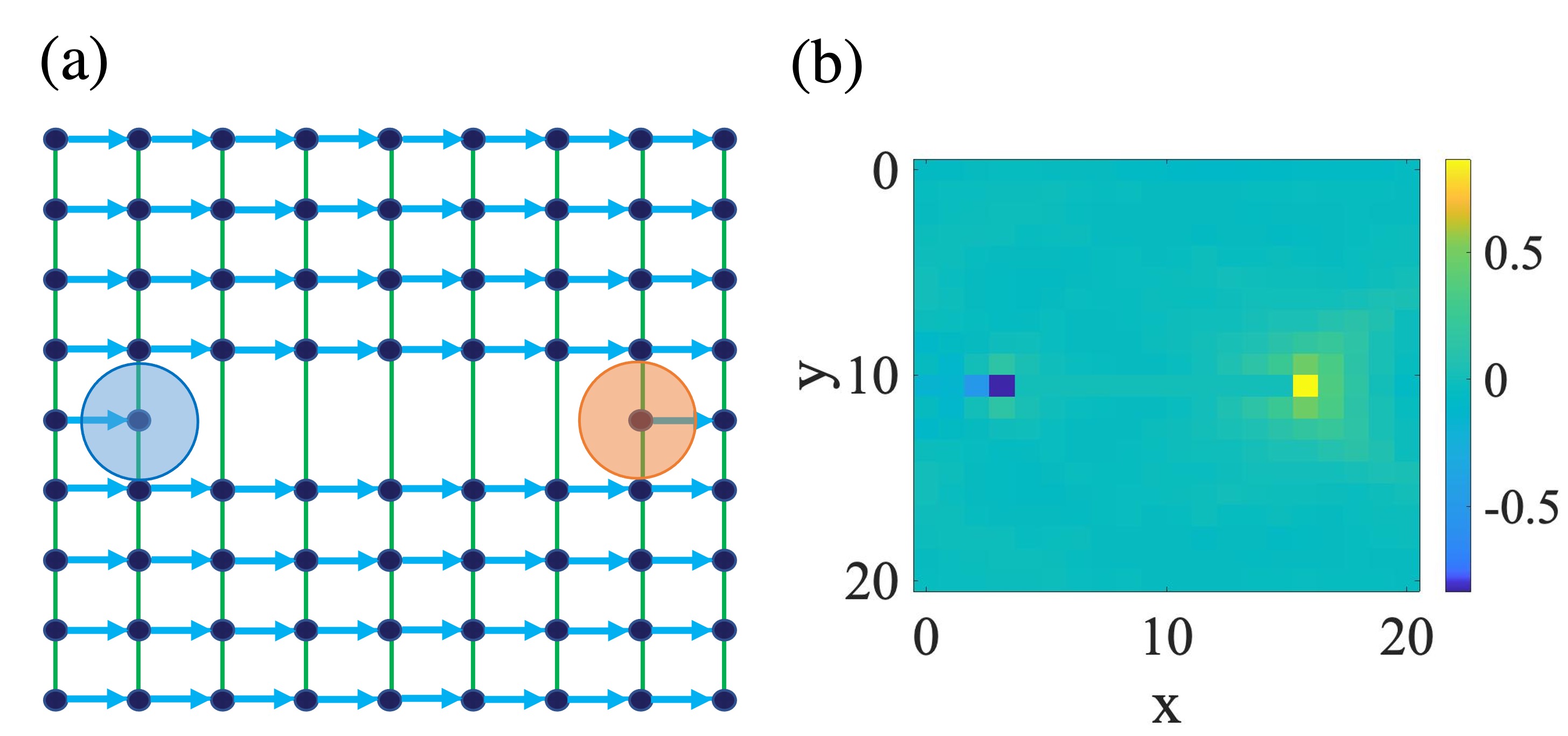}
    \caption{\label{Fig: Dislocation}\textbf{Dislocation induced skin effect and anti-skin effect in a two-dimensional non-Hermitian system.} Panel (a) shows the square lattice where two dislocation centres have been introduced, shown by the blue and orange highlighted circles. The hopping in the $x$-direction is non-reciprocal -- $t_x+\delta_x$ in the rightward direction (shown by the blue arrows), and $t_x-\delta_x$ in the leftward direction. The green lines coupling the lattice sites in the $y$ direction denote isotropic hopping strength $t_y$. Panel (b) shows the occurrence of the skin effect and anti-skin effect where the colours denote values of $\log \rho_r$. Here, the system is a $20\times 20$ square lattice with $t_x=1,t_y=1$ and $\delta_x=0.5$.}
\end{figure*}

In this context, we would also like to draw the reader's attention to a recently discovered new type of skin effect. Here, crystal defects in the form of `dislocations' can act as terminal points within the lattice and cause the localization of eigenstates in their vicinity. This has been termed the dislocation non-Hermitian skin effect (DNHSE)~\cite{schindler2021dislocation,panigrahi2022non,jing2022topological}. Bhargava \textit{et al.}~\cite{bhargava2021non} recently made an interesting discovery that in a two-dimensional system with two dislocation sites, there not only arises a dislocation-induced skin effect at one site, but there occurs an `anti-skin effect' at the other site. Further, they identify a topological invariant -- the $\mathbb{Z}_2$ Hopf index to characterize the DNHSE. For illustration, here we adopt a model similar to theirs in order to demonstrate this skin effect and anti-skin effect in a system with dislocations. We consider a square lattice model as shown in Fig.~\ref{Fig: Dislocation}(a). The Hamiltonian of the model is given as follows

\begin{equation}
    H= \sum_r (t_x+\delta_x) c_{r+x}^\dagger c_r +(t_x-\delta_x) c_{r-x}^\dagger c_r + t_y(c_{r+y}^\dagger c_r + c_{r-y}^\dagger c_r), \label{eq:dis_Ham}
\end{equation}

where, $r$ denotes the position index. Hopping in the $x$-direction is non-reciprocal -- $t_x+\delta_x$ is the rightward hopping, shown by blue arrows in Fig.~\ref{Fig: Dislocation}(a), while the leftward hopping is $t_x-\delta_x$. In the $y$-direction, the hopping is isotropic and takes the value $t_y$. In this lattice, we introduce two dislocations, i.e., we select two lattice sites (or dislocation centres), as shown in Fig.~\ref{Fig: Dislocation}(a) and remove the row of atoms in between. Further, we reintroduce $t_y$ between the atoms of the existing adjacent rows. This procedure is called the `cut and glue method'. Next, to check the effect of the dislocations on the localization behaviour of the system, we impose PBC along both $x$ and $y$ directions, such that the dislocation centres are the only terminations in the system. We plot the quantity $\text{log } \rho_r$ in Fig.~\ref{Fig: Dislocation}(b), where $\rho_r=\sum_n |\langle r|\psi_n\rangle|^2$ and $|\psi_n\rangle$ is the $n$-th right eigenvector. The plot of $\text{log } \rho_r$ shows the occurrence of the skin effect and anti-skin effect at the right and left dislocation centres, respectively.

The NHSE, which is a uniquely non-Hermitian phenomenon, leads to immensely varying eigenstate behaviours depending on the boundary conditions imposed on the system. This leads to the failure of the well-known bulk-boundary correspondence in non-Hermitian systems, which we discuss further in the next section.

\section{Broken Bulk-Boundary Correspondence and\\ the Generalized Brillouin Zone} \label{generalized Brillouin Zone}

In general, for topological phases of Hermitian systems, there exists a topological invariant which depends on the bulk properties of the system, which can predict the number of zero energy boundary states the system will exhibit~\cite{hasan2010colloquium}. The existence of edge states solely depends on the bulk of the system. This makes them robust to perturbations and hence topological in nature. However, the presence of non-Hermiticity, leads to a failure of the conventional bulk boundary correspondence~\cite{helbig2020generalized,xiong2018does,jin2019bulk,kunst2019non,kawabata2020non,zhu2020photonic}. The NHSE, which is accompanied by the drastic localization of all eigenstates, demonstrates the strikingly different behaviour of the system at the bulk and the boundary. This extreme sensitivity of non-Hermitian systems to their boundary conditions leads to the broken bulk boundary correspondence, which posed a considerable challenge. For instance, winding numbers calculated for the bulk cannot predict the number of boundary states as there is a significant difference between the PBC and the OBC spectra.

In order to resolve this problem and establish a generalized bulk boundary correspondence, the notion of a generalized Brillouin zone (GBZ) was proposed by Yao and Wang~\cite{yao2018edge}. Here, the usual Bloch phase factor $e^{ik}$ is replaced by a non-Bloch phase factor $\beta= r e^{ik}$ for the system under PBC, where the system parameters determine $r$. This implies that the wavevector acquires an extra imaginary factor such that $k \rightarrow k -i \ln r$. Here $\beta$ can take values in a non-unit circle $\abs{\beta}=r$, which defines the GBZ. The non-Bloch Hamiltonian, $H(\beta)$, can now be used to calculate a winding number $W_{\textrm{GBZ}}$ over the GBZ, and $2 W_\textrm{GBZ}$ determines the number of robust zero energy modes at the boundaries of the system under OBC. This generalized Bloch band theory helps substantiate the generalized bulk boundary correspondence for non-Hermitian systems, which was further demonstrated by Yokomizo and Murakami in one-dimensional systems \cite{yokomizo2019non}. Moreover, to explain the occurrence of the NHSE under OBC, a similar effective Hamiltonian $H(\beta)$ can be written in terms of the GBZ, where $\beta$ is determined by the system parameters. To illustrate, for a one-dimensional tight-binding model on a bipartite lattice, according to Ref.~\cite{wang2019non}, the eigenstates on lattice sites $A$ and $B$ can be written as $(\psi_{An}, \psi_{Bn})= \beta^n (\psi_A,\psi_B)$, where $n$ denotes the $n$-{th} lattice site. When $|\beta|=1$, all the eigenstates are extended. However, $|\beta| \neq 1$ implies an NHSE, where $\beta<1$ and $\beta>1$ signify localization of the states at opposite edges, establishing a connection between the boundary-behaviour and the bulk Hamiltonian under OBC.

As a concrete example, let us consider the generalized non-Hermitian SSH model (as shown in Fig. \ref{SSH-figure}(a)) and construct the GBZ closely following the methods specified in Refs.~\cite{yao2018edge,yokomizo2019non}. The real space Schr{\"o}dinger equations for the SSH model can be written as

\begin{equation}
\begin{split}
    t_2 \psi_{B_{n-1}} +  t_3 \psi_{B_{n+1}} + (t_1+\delta) \psi_{B_{n}}+ i \gamma \psi_{A_{n}}= E \psi_{A_{n}},\\
    t_2 \psi_{A_{n+1}} +  t_3 \psi_{A_{n-1}} + (t_1-\delta) \psi_{A_{n}}- i \gamma \psi_{B_{n}}= E \psi_{B_{n}}.
\end{split}  
\end{equation}

 Now, we consider the ansatz $(\psi_{A_n},\psi_{B_n})=\beta^n (\psi_A,\psi_B$). Then, the above equations become

\begin{equation}
\begin{split}
    t_2 \beta^{-1} \psi_{B} +  t_3 \beta \psi_{B} + (t_1+\delta) \psi_{B}+ i \gamma \psi_{A}= E \psi_{A},\\
    t_2 \beta \psi_{A} +  t_3 \beta^{-1} \psi_{A} + (t_1-\delta) \psi_{A}- i \gamma \psi_{B}= E \psi_{B}.
\end{split}
\end{equation}

Multiplying the equations and cancelling $\psi_A \psi_B$ from both sides gives us

\begin{equation}
    (t_2 \beta^{-1} + t_3 \beta +t_1 +\delta) (t_2 \beta +t_3 \beta^{-1} +t_1 -\delta)=E^2 +\gamma^2.
\end{equation}

From this, we obtain a quadratic equation in $\beta$ as follows

\begin{equation}
\begin{split}
    t_2 t_3 \beta^4 +(t_1 t_3 +t_1 t_2 - \delta(t_3-t_2))\beta^3 + (t_2^2+t_3^2+t_1^2-\delta^2\\-\gamma^2+E^2)\beta^2+ (t_1t_2+t_1t_3+\delta(t_3-t_2))\beta+t_2t_3 =0.
\end{split}
\end{equation}

\begin{figure*}
    \centering
    \includegraphics[width=0.95\textwidth]{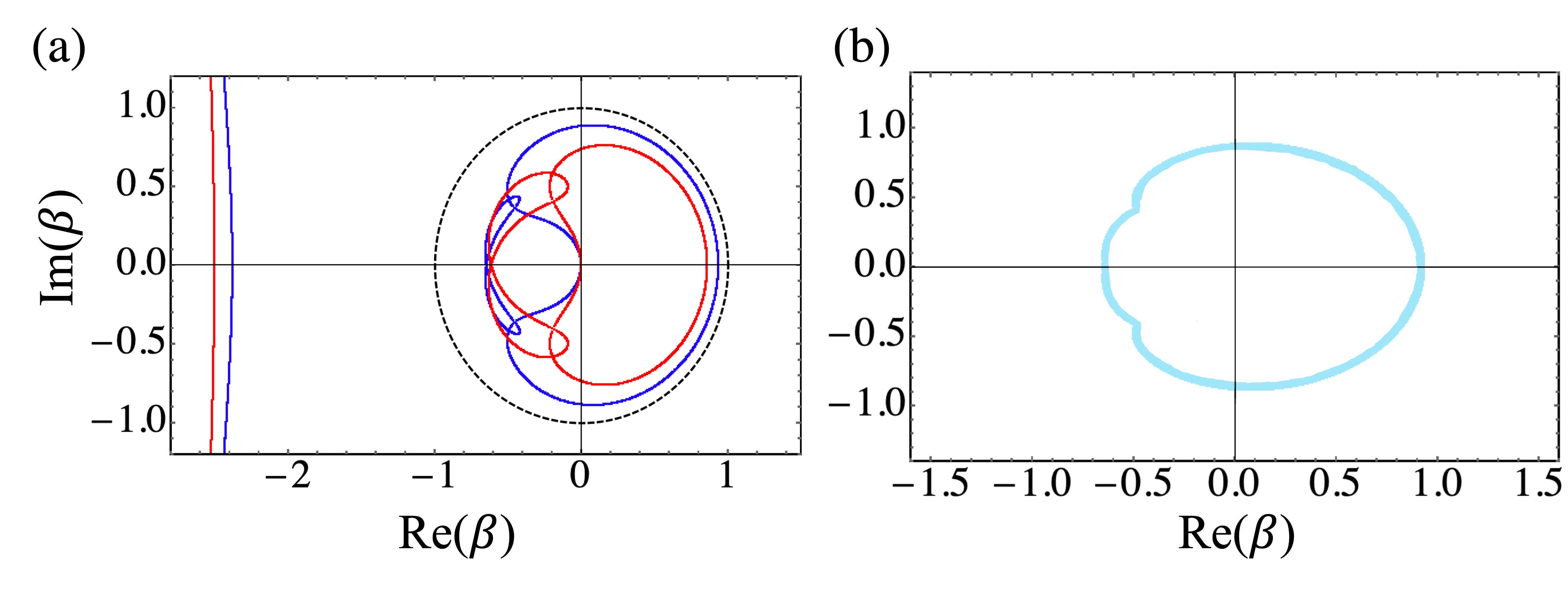}
    \caption{\label{Fig: GBZ}\textbf{The generalized Brillouin zone for the non-Hermitian SSH model.} Panel (a) shows the auxiliary GBZ whose subset is the GBZ shown in panel (b). (a) The black dashed lines correspond to the Hermitian case of the SSH model where the GBZ is a unit circle. The blue curves show $C_\beta$ where $t_3=1/3$ and the non-reciprocal hopping $\delta=1/4$. The red curves correspond to $t_3=2/5$ and $\delta=2/3$. $C_\beta$ in the presence of $t_3$ and non-Hermiticity distorts from a circle. The $C_\beta$ trajectories have been plotted such that $|\beta_i|=|\beta_j|$ ($i\neq j$). (b) The GBZ for the case corresponding to the blue curves ($t_3=1/3,\delta=1/4$). Here, we plot the trajectories of $\beta_2$ and $\beta_3$, such that, $\abs{\beta_1}\le\abs{\beta_2}=\abs{\beta_3}\le \abs{\beta_4}$. For all plots the values of other parameters are chosen to be $t_1=1.3$, $t_2=1.5$ and $\gamma=0.5$.}
\end{figure*}

In order to obtain the trajectory of the GBZ ($C_\beta$), $\beta$ must satisfy the condition $|\beta_i|=|\beta_j|$ for a pair $i \neq j$. The above equation is of the form $f(\beta)=-E^2$. We consider $\beta'=\beta e^{ik}$ such that $k \in \mathbb{R}$, which will also satisfy $f(\beta')=-E^2$. Taking the difference of $f(\beta)-f(\beta')$, we arrive at a quadratic equation for $\beta(k)$ of the form 

\begin{equation}
\begin{split}
    t_2 t_3 \beta^2 (1-e^{2ik}) +(t_1 t_3 +t_1 t_2 - \delta(t_3-t_2)) \beta (1-e^{ik}) +(t_1t_2 \\+t_1t_3 +\delta(t_3-t_2)) \beta^{-1} (1-e^{-ik})+ t_2t_3 \beta^{-2} (1-e^{-2ik})=0,
\end{split}
\end{equation}

Plotting $\beta(k)$ in the complex plane for $k \in [0,2\pi]$, we obtain the trajectory $C_\beta$ shown in Fig.~\ref{Fig: GBZ}(a). When the non-Hermiticity $\delta=0$, $C_\beta$ becomes a unit circle ($|\beta|=1$) shown by the black dashed lines. The blue and red curves show $C_\beta$ for different values of $t_3$ and $\delta$. The GBZ corresponding to the blue curves is shown in Fig.~\ref{Fig: GBZ}(b) where $\beta_2$ and $\beta_3$ have been plotted such that the condition $\abs{\beta_1}\le\abs{\beta_2}=\abs{\beta_3}\le \abs{\beta_4}$ is satisfied. $|\beta|\neq 1$ implies the existence of NHSE in the system under OBC.\\
Next, we take a look at the effects of disorder in non-Hermitian systems. Later, in the section, we specifically address how disorder can have interesting consequences in context of the skin effect.

\section{Disorder Effects in Non-Hermitian Systems}

Disorder, being ubiquitous in nature, has gained a lot of interest~\cite{kim2021disorder,zeng2020winding,longhi2021spectral,sarkar2022non,gong2018topological}, including in the context of non-Hermitian settings. The interplay of disorder and non-Hermiticity has led to the discovery of several new physical phenomena, some entirely unique to non-Hermitian systems~\cite{zhang2020non,longhi2021spectral,claes2021skin,liu2021real,liu2022modified,zhang2022bulk}. The study of disorder leads to the requirement of quantities which are `self-averaging' under disorder, i.e., good indicators which should be independent of the exact disorder configuration but should depend only on the strength and the nature of the disorder introduced. Mondragon-Shem \textit{et al}.~\cite{mondragon2014topological} introduced a real-space winding number which is self-averaging, robust and remains quantized even at strong disorder values. Though this was proposed for Hermitian systems, this formalism was later extended to disordered non-Hermitian cases as well. A change in the winding number at a critical disorder strength leads to a phase transition where they have analytically proven that the localization length diverges. Localization transitions induced by non-Hermitian disorder was further elucidated in one-~\cite{wang2021anderson,zhang2021non}, two-~\cite{claes2020disorder,tzortzakakis2020non}, and three-dimensional systems~\cite{huang2020anderson,luo2021universality,huang2020spectral}, as well as in non-Hermitian quasi-periodic crystals~\cite{jiang2019interplay,liu2021localization}. Another exciting report by Hamazaki \textit{et al.}~\cite{hamazaki2019non} showed that in a time-reversal symmetric non-Hermitian system disorder-induced many-body localization can lead to a complex to real transition of the energy eigenspectrum. Refs.~\cite{luo2022unifying,luo2021universality} explored the behaviour of critical exponents and report on how non-Hermiticity modifies the universality classes of Anderson transitions. Another interesting recent discovery is the `non-Hermitian topological Anderson insulator' -- a normal insulator which transforms into a topological insulator under the effect of disorder and is characterized by a change in the topological winding number and occurrence of zero energy edge modes~\cite{luo2019non,zhang2020non,liu2020topological,tang2020topological}. Further, in Ref.~\cite{zyuzin2018flat}, the authors showed that disorder in a type-II non-Hermitian Weyl semimetal can exhibit a flat band. Longhi~\cite{longhi2021spectral} studied the role of different kinds of disorder in the Hatano-Nelson model by investigating the deformations in the radius of the complex energy spectrum. He discovered three disorder-induced phases in the model under both PBC and OBC, which was further confirmed by Sarkar \textit{et al.} for the non-Hermitian SSH model~\cite{sarkar2022interplay}. Claes and Hughes~\cite{claes2021skin} discovered the fascinating `non-Hermitian Anderson skin effect' (NHASE) in the Hatano-Nelson model where a non-Hermitian system initially exhibiting no skin effect develops an NHSE under the effect of disorder which is accompanied by a transition in the winding number to non-zero values. Sarkar \textit{et al.}~\cite{sarkar2022interplay} discovered a similar NHASE in the non-Hermitian SSH model and investigated the role of different non-Hermitian symmetries in this context. Interestingly, the NHASE occurs under disorder in symmetry classes A, AIII and D$^\dagger$. Notably, however, only when all the symmetries of the system are broken does the real space winding number show a direct correspondence to the NHASE.

There have also been recent studies of disorder in non-Hermitian systems using field theoretical tools~\cite{moustaj2022field} and transfer matrix calculations~\cite{luo2021transfer}. Wanjura \textit{et al.}~\cite{wanjura2021correspondence} discovered that non-Hermitian systems can exhibit directional amplification even in the presence of complex (local or non-local) disorder, as long as the size of the point gap in the spectrum remains larger than the maximum disorder introduced into the system. Further, Ref.~\cite{kim2021disorder} reported the effect of disorder on higher-order NHSE. Recently, Okuma and Sato~\cite{okuma2021non} have generalized the concept of NHSE to Hermitian disordered systems via the Green's function approach, where a Hermitian Hamiltonian modified by its self-energy effectively acts as a non-Hermitian one.

\begin{figure*}
    \centering
    \includegraphics[width=0.8\textwidth]{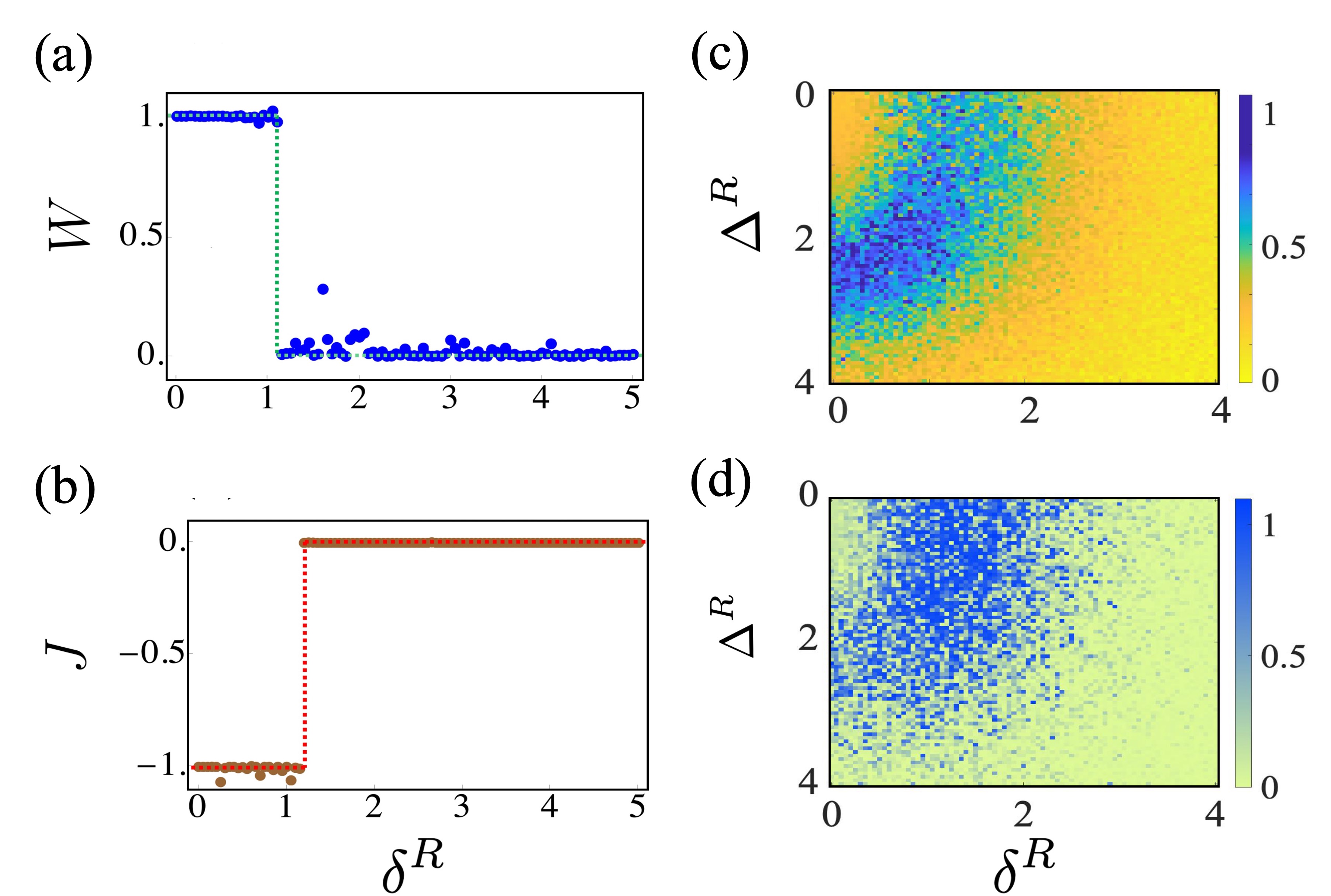}
    \caption{\label{Fig: Disorder}\textbf{Equivalence between the current and the winding number under disorder and disorder induced NHASE.} (a) The winding number, $W$, as a function of disorder under PBC. (b) The chiral current as a function of the same hopping disorder. (a) and (b) show that the winding number and chiral current are equivalent, both showing quantized and robust behaviour under disorder. Panel (c) shows the NHASE where a non-Hermitian skin effect develops in the system under disorder when one starts from a set of parameters without skin effect. Here the colour plot denotes the edge density as a function of both hopping disorder and onsite disorder. Panel (d) shows the real space winding number for the system with all broken symmetries demonstrating a correspondence with the edge density under the broken symmetry condition. For (a) and (b) $t_1=1.0$, $t_2 =1.0$, and $\delta=0.3$, whereas, for (c) and (d) $t_1=1.1$, $t_2 =1.0$, and $\delta=2.1$. For all plots, we use 400 lattice sites and $\gamma =0$.}
\end{figure*}

As an illustration of the interesting effects of disorder (adapted from our previous work~\cite{sarkar2022interplay}), we consider the non-Hermitian SSH model with nearest neighbour couplings, non-reciprocal intracell hopping and complex onsite potentials. Disorder is introduced into the system via the non-reciprocity parameter ($\delta$) and the on-site energies ($\gamma$). Each non-reciprocal hopping in the real space becomes $\delta_i = \delta +\delta^R_i$ where each $\delta^R_i \in \delta^R[-1,1]$. Here, $\delta$ is the mean value about which disorder of strength $\delta^R$ is added. A similar set of random values is added to the on-site gain and loss terms $\gamma_i = \gamma + \Delta^R_i$ where $\Delta^R_i \in \Delta^R[-1,1]$ and the disorder strength is denoted by $\Delta^R$. We compute the real space winding number and the current (given in Eq.~\ref{eq:chiralcurrent}) as a function of increasing disorder. In order to calculate the real space winding number we define a base energy, $E_b$ and construct the matrix $H-E_b I$. On performing a singular value decomposition, we obtain the form $H-E_b I=M S N^{\dagger}$, where $S$ is a diagonal matrix with eigenvalues along the diagonal. Then, one can define $Q=M N^{\dagger}$ and $P=N S N^{\dagger}$, such that $H-E_b I=QP$. From this polar decomposed form of the Hamiltonian we can write

\begin{equation}
    W= \frac{1}{L'} \mathrm{Tr'} (Q^{\dagger} [Q,X]). \label{eq:windingreal}
\end{equation}

Here $X$ is the position operator. The trace is taken only over the bulk of the system so a sufficient number of lattice sites is eliminated from both edges. The effective bulk length is $L'=L-2l$, where $L$ is the total length of the lattice while $l$ is a cutoff length.

As shown in Fig.~\ref{Fig: Disorder}, plotting $W$ and $J$ with respect to the disorder strength values, we find that there is an exact equivalence between the winding number and the current. Interestingly, the current shows a robust and quantized behaviour retaining its clean system value till it makes a sharp transition and drops to zero exactly where the winding number too becomes trivial. Fig.~\ref{Fig: Disorder} (a) and (b) are shown for disorder in $\delta$ however, qualitatively the same behaviour is obtained for disorder in the onsite energies \cite{sarkar2022non}. In Fig.~\ref{Fig: Disorder} (c), we illustrate the NHASE in the same system. Starting from a region with no skin effect, increasing either kind of disorder ($\delta^R$ or $\Delta^R$) or both, the system develops a skin effect (shown by a large increase in edge density) dubbed the NHASE. The edge density is calculated as the local density of states at the edge of the system given by $\frac{\sum_{x_i=1}^{x_E} \abs{\psi_{\alpha}(x_i)}^2}{\sum_{x_i} \abs{\psi_{\alpha}(x_i)}^2}$, where $x_E$ denotes the width of the edge. As mentioned in the previous discussion, the NHASE occurs for several symmetry classes. However, when all the symmetries of the system are broken, the real space winding number shows an exact equivalence to the NHASE (as shown in Fig.~\ref{Fig: Disorder}(d)]).

\section{Non-Hermitian linear response theory}

With growing progress in the understanding of non-Hermitian systems, it has become important to understand how they respond to external perturbations. This has necessitated the development of new paradigms, which we discuss next. Linear response theory has been widely used to study the response of a system under the influence of an external source~\cite{fetter2012quantum,pines2018theory,giuliani2005quantum}. When the effect of this external source acts as a small perturbation to a quantum mechanical system, the change in the expectation value of any operator is assumed to be linear in perturbation

\begin{equation}
   \delta \langle  O(t) \rangle=\int d t'\chi(t,t') \phi_{ext}(t'),
\end{equation}

where $\chi(t,t')$ is known as the response function, $\delta \langle  O(t) \rangle$ is the change in the expectation value of an operator and $\phi_{ext}(t)$ is the external perturbation acting on the system. With the help of the linear response theory, one can study electrical transport through the Kubo formula and fluctuations through the celebrated fluctuation-dissipation theorem on a system and analyze the response of systems when coupled with suitable experimental probes~\cite{shi1998effects,tokuno2011spectroscopy}. However, these concepts are well-established only in the Hermitian realm. Non-Hermitian physics requires a modified version of the linear response theory.

The first attempt at developing a linear response theory in the context of non-Hermitian systems was taken by Pan \textit{et al.}~\cite{pan2020non}. In their work, they considered a Hermitian system with a non-Hermitian perturbation. The linear response of the system with non-Hermiticity was used to extract information about the parent equilibrium state. This served as the foundation of non-Hermitian linear response theory. The success of this theory was multi-fold. They discovered that the response of the `real-time' dynamics of the momentum distribution of a quantum state under dissipation is governed by the single-particle `real-time' spectral function at equilibrium. This remarkable observation made it possible to distinguish between quantum states with well-defined quasi-particles and the critical states without well-defined quasi-particles. Furthermore, by re-investigating the data obtained in an experiment dealing with the dissipative Bose-Hubbard model in a cold-atom setup~\cite{bouganne2020anomalous}, they inferred that performing a single measurement is sufficient to obtain information about the height and width of the momentum distribution of the system. Finally, using these observations, they calculated the critical exponent for the superfluid to Mott insulator phase transition for the first time.

After Pan \textit{et al.}, non-Hermitian linear response theory was further generalized by Striclet \textit{et al.}~\cite{sticlet2022kubo}. They considered a non-Hermitian perturbation to an underlying non-Hermitian system. We briefly discuss this formalism here. In this framework, the change in an operator $A(t)$ to linear order for such a system is given by,

\begin{equation}
    \delta\langle A(t) \rangle=\frac{\mathrm{tr}[\delta\rho_{I}(t)A_{I}(t)]}{\mathrm{tr}[\rho_{0}(t)]}-\langle A(t) \rangle _{0}\frac{\mathrm{tr}[\delta\rho(t)]}{\mathrm{tr}[\rho_{0}(t)]},
\end{equation}

where $\rho_{0}(t)$ is the density matrix of the unperturbed system, the contribution from $\delta\rho(t)$ is owed to the perturbation, $\langle A(t) \rangle _{0}$ is the expectation value determined in the unperturbed system and $H_{0}$ is the unperturbed non-Hermitian Hamiltonian. Following the interaction picture representation, one has $\delta\rho_{I}(t)=\text{e}^{i H_{0}t/\hbar}\delta\rho(t)\text{e}^{-i H_{0}^{\dagger}t/\hbar}$ and $A_I(t)=\text{e}^{i H_{0}^{\dagger}t/\hbar}A\text{e}^{-i H_{0}t/\hbar}$. The terms in the denominator are required for proper normalization as the unperturbed Hamiltonian is non-Hermitian. Additionally, the second term appears because of the non-unitary dynamics of the non-Hermitian picture as a norm correction term; hence is absent in Hermitian linear response theory. The generalized response function is obtained to be

\begin{equation}
    \chi_{AB}(t,t')=-\frac{i}{\hbar}\theta(t-t')\mathrm{tr} \Biggl\{ \textbf{[} [A(t-t'),B]_{\sim}-\langle A(t) \rangle_{0}[\text{e}^{i H_{0}^{\dagger}(t-t')/\hbar}\text{e}^{-i H_{0}(t-t')/\hbar},B]_{\sim}\textbf{]}.\frac{\rho_{0}(t')}{\mathrm{tr}[\rho_{0}(t)]}\Biggr\},\label{eq:NH_response}
\end{equation}

where $B$ is the operator associated with the external time-dependent perturbation, $[X,Y]_{\sim}=XY-Y^{\dagger}X$ is the modified commutator. The first term in the response function is the modified Kubo formula for non-Hermitian cases and second term, as mentioned before, is the norm correction term. Using Eq.~\eqref{eq:NH_response}, the standard Kubo formula can be recovered by setting $H_0=H_0^{\dagger}$ and $B=B^{\dagger}$. Furthermore, one can retrieve the earlier results by Pan \textit{et al.}~\cite{pan2020non}, i.e., the effect of a non-Hermitian perturbation on a Hermitian system by setting $H_0=H_0^{\dagger}$ but $B\neq B^{\dagger}$. 

Using this theory, Striclet \textit{et al.}~\cite{sticlet2022kubo} showed that a one-dimensional non-Hermitian relativistic Dirac model with a tachyon phase has a finite conductivity, similar to the two-dimensional graphene sheet. The imaginary mass of the particles results in the non-Hermiticity in the model. Away from the EPs, the expectation value of the current in this model determined using linear response theory and those from the numerical solution of the Schr$\ddot{\text{o}}$dinger equation correspond almost perfectly. Further, this non-Hermitian linear response theory allows us to determine the unequal-time anti-commutator of observables, making it possible to observe the fluctuation-dissipation relations in quantum systems~\cite{geier2022non}.

\section{Transport signatures of non-Hermitian systems} \label{transport}

As we have learned from Hermitian topological phases, many key topological features manifest in the transport properties of the system. In the last few years, extensive research has been undertaken to understand how the presence of non-Hermiticity modifies the transport phenomena in a non-Hermitian system~\cite{chen2018hall,yao2018non,longhi2015robust}. Here, we summarize the recent findings about an important transport signature, namely the Hall conductance of non-Hermitian topological systems.

For the Hermitian two-dimensional Chern insulator, the Chern number associated with the filled bands, determines the number of edge modes of the gapped insulator. The Hall conductance is proportional to the sum of the Chern numbers of all the filled bands given by the Thouless-Kohmoto-Nightingale-den Nijs (TKNN) formula~\cite{thouless1982quantized}. When the systems come into contact with the environment, the effective Hamiltonian describing the system becomes non-Hermitian. The imaginary term in the Hamiltonian broadens the density-of-states in the band gap in the complex energy plane. If the minimum separation between the real part of the eigenenergies is much larger than the imaginary part, the non-Hermitian bands can be considered to be ``gapped". 

Let us now look at the Hall conductance for a generic non-Hermitian two-band Hamiltonian of the form

\begin{equation}
    H(\textbf{k})=(d^{0}_{\textbf{k}}-i\Gamma^{0}_{\textbf{k}})\sigma_{0}+(\textbf{d}_{\textbf{k}}-i\boldsymbol{\Gamma}_{\textbf{k}}).\boldsymbol\sigma_{\textbf{k}},
\end{equation}

where $\textbf{d}_{\textbf{k}}=(d_{\textbf{k}}^{x},d_{\textbf{k}}^{y},d_{\textbf{k}}^{z})$ and $\boldsymbol{\Gamma}_{\textbf{k}}=(\Gamma_{\textbf{k}}^{x},\Gamma_{\textbf{k}}^{y},\Gamma_{\textbf{k}}^{z})$ are the three component vectors. The current-current correlation is given by

\begin{equation}
    K_{\alpha\beta}(\omega)=\sum_{\textbf{k}}\int d\epsilon d\epsilon'\frac{n_{F}(\epsilon')-n_{F}(\epsilon)}{\epsilon'-\epsilon+\omega+i0^{+}}\text{tr}(\hat{J_{\alpha}}A(\epsilon)\hat{J_{\beta}}A(\epsilon')),
\end{equation}

where $n_{F}(\epsilon)$ is the Fermi distribution function at temperature $T$, $\hat{J}_{\alpha,\beta}=\frac{\partial \text{Re}H(\textbf{k})}{\partial k_{\alpha,\beta}}$ is the current operator, $ A(\epsilon)=\text{Imtr}\hat{G}^{R}$ is the spectral function, and $\hat{G}^{R}$ is the retarded Green's function. Then using the linear response theory, the Kubo formula for the Hall conductance of the system is defined as

\begin{equation}
    \sigma_{\alpha\beta}=\lim_{\omega\rightarrow 0}\frac{i}{\omega+i0^{+}}[K_{\alpha\beta}(\omega)-K_{\alpha\beta}(0)].
\end{equation}

The zero temperature limit of $\sigma_{\alpha\beta}$ can be obtained as

\begin{equation}
    \sigma_{xy}=\sum_{\textbf{k}}\frac{\Omega_{xy}(\textbf{k})+\Omega^{*}_{xy}(\textbf{k})}{2}\times\left(\frac{2}{\pi}\arctan{\frac{\text{Re}~\epsilon(\textbf{k})}{\text{Im}~\epsilon(\textbf{k})}}\right).\label{eq:Hall_f}
\end{equation}

Here, $\Omega_{xy}(\textbf{k})=\textbf{h}_{\textbf{k}}.(\partial_{k_x}\textbf{d}_{\textbf{k}}\times \partial_{k_y}\textbf{d}_{\textbf{k}})/\epsilon(\textbf{k})^3$, $\textbf{h}(\textbf{k})=(\textbf{d}_{\textbf{k}}-i\boldsymbol{\Gamma}_{\textbf{k}})$ and $\epsilon(\textbf{k})$ is the eigenenergy of the Hamiltonian. The presence of the second term is a manifestation of the non-Hermiticity and it can take any value between $[0,1]$ depending on the strength of the non-Hermitian terms present in the system. Eq.~\eqref{eq:Hall_f} is the non-Hermitian version of the TKNN formula as obtained in Ref~\cite{chen2018hall}.

As a concrete illustration, we use the previously discussed Chern insulator model (please see Section~\ref{models} C) to analyze the Hall conductance. The Hamiltonian is given by

\begin{equation}
    H(\textbf{k})=(\sin{k_x}+i\gamma (\textbf{k}))\sigma_x+(\sin{k_y}+i\gamma (\textbf{k}))\sigma_y+(m+\cos{k_x}+\cos{k_y})\sigma_z.\label{eqn:NHCmodel}
\end{equation}

\begin{figure}
    \centering
    \includegraphics[width=0.95\textwidth]{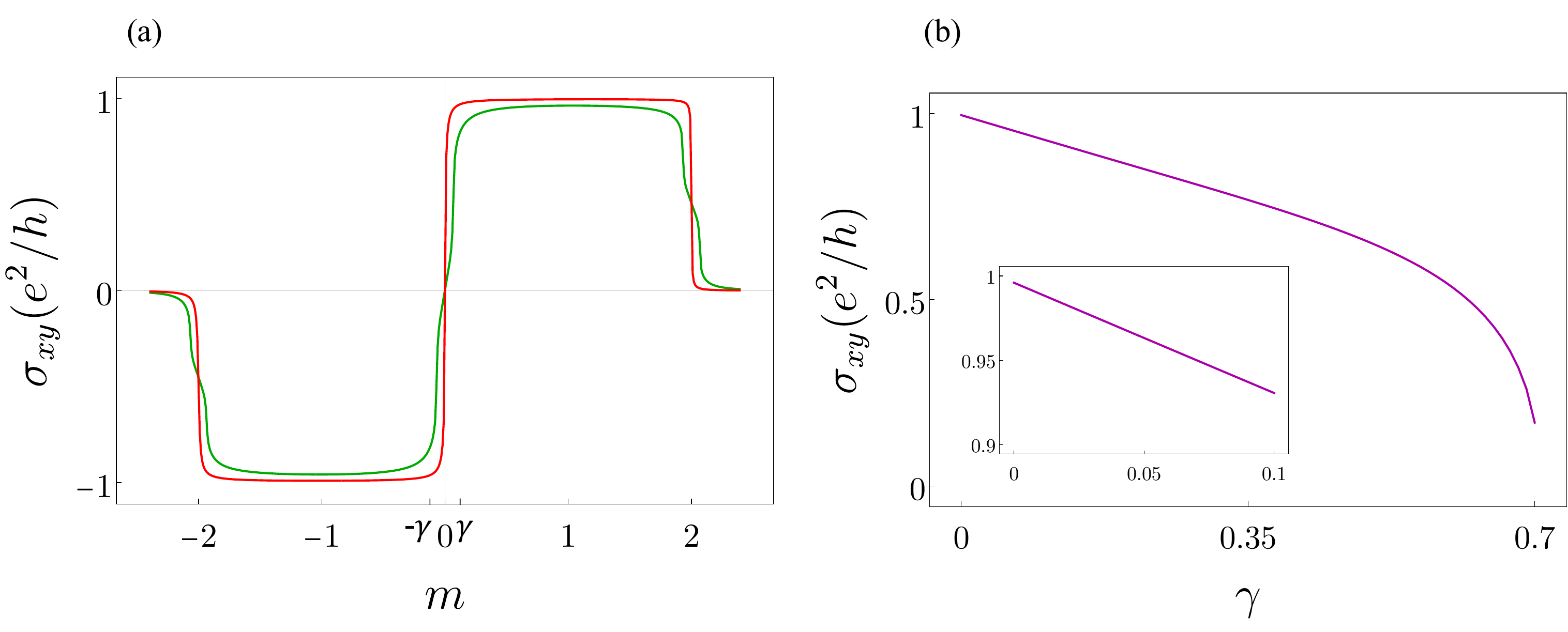}
    \caption{\textbf{Hall conductance of a non-Hermitian Chern insulator.} (a) The Hall conductance of the non-Hermitian Chern insulator (green curve) is shown along with the Hall conductance of the model in absence of the non-Hermiticity (red curve) as a function of $m$. Here, $\gamma$ is taken to be momentum independent and has a numerical value of $0.05$. Panel (b) shows the Hall conductance as a function of the momentum independent non-Hermiticity parameter $\gamma$ when $m=1$. The inset shows the linear scaling at small values of $\gamma$.} 
    \label{fig:Hall}
\end{figure}

When $\gamma\rightarrow0$, i.e., in the Hermitian limit, this Hamiltonian describes a Hermitian Chern insulator. For $-1<m<0$, $\sigma_H=-1$ and for $0<m<1$, $\sigma_H=1$. For other values of $m$, we have $\sigma_H=0$. Here $\gamma$ can in general, be dependent on momentum, and $|\gamma(\textbf{k})|$ is always less than unity, such that the imaginary part of the eigenenergies always lie in the negative half of the complex energy plane.
Strikingly, even if the non-Hermitian Chern number remains quantized for a ``gapped" non-Hermitian insulator, the Hall conductance is no longer a quantized quantity. In Fig.~\ref{fig:Hall}(a) the Hall conductance for this non-Hermitian model is plotted as a function of $m$ considering the imaginary terms to be momentum independent. Whenever $m$ lies in between $\pm\gamma$, the deviation from quantization is more substantial. This deviation increases and the Hall conductance decreases with the increasing strength of $\gamma$, as shown in Fig.~\ref{fig:Hall}(b). The inset in Figure~\ref{fig:Hall}(b) shows that the Hall conductance for low values of $\gamma$ varies linearly. 

The non-universal deviation of Hall conductance of a lossy Chern insulator from quantization is due to the broadening of the spectral function gap and the decay term present in the energy of the edge modes~\cite{chen2018hall, philip2018loss, wang2022hall}. In other words, the probability that the carriers will make it to the other end during transport is less than unity. With increasing strength of the imaginary term in the Hamiltonian, the Hall conductance monotonically decreases. Further, it was also noted that the bulk contribution to the Hall conductance of a non-Hermitian Hamiltonian is non-zero~\cite{chen2018hall}. Additionally, a new contribution to the Hall conductance emerges when the system couples to the environment in such a way that the system experiences compensated momentum-independent gain and loss~\cite{groenendijk2021universal}. This universal non-analytic contribution follows a $3/2$-power law relation with the system parameters (mass term, chemical potential, gain and loss parameter). Very recently, a few theoretical works on Hall conductance in three-dimensional non-Hermitian time-reversal-breaking systems have also appeared~\cite{he2020floquet,wu2022non}. An in-depth understanding of the transport properties of such systems will be worth pursuing.

\section{Non-equilibrium steady states and many-body phases in non-Hermitian systems} \label{many-body}

\begin{figure*}
\includegraphics[scale=0.45]{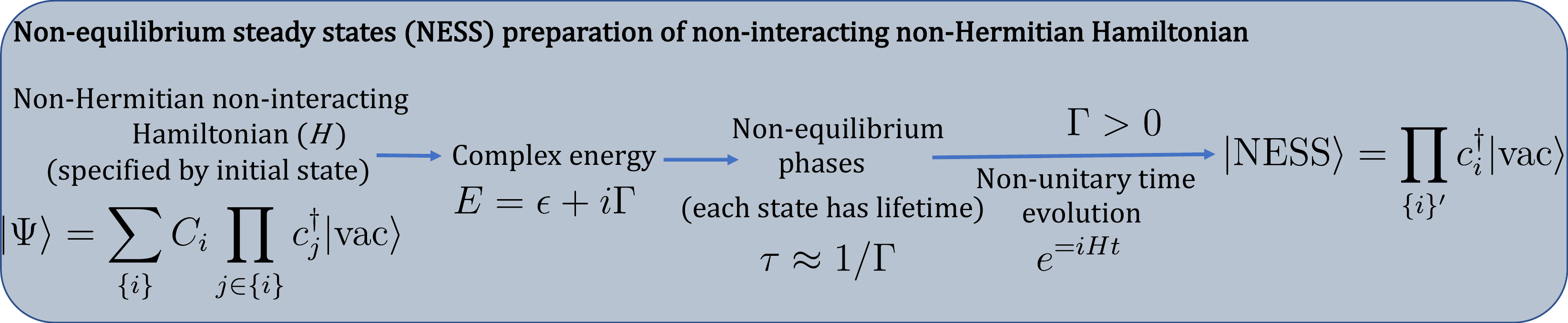}
  \caption{\textbf{Preparation of NESS for a non-Hermitian non-interacting model.} We consider a non-Hermitian non-interacting Hamiltonian specified by initial state $|\Psi\rangle$ (see Ref.~\cite{lieu2019non}). Here ${i}$ denotes all possible combinations of occupied states and $C_{i}$ is the associated weight. The imaginary part of the eigenenergy defines a lifetime for each single particle eigenstate. Any generic many-body state under unitary time evolution will eventually evolve into the many-body steady state ($\text{NESS}$) with the largest value of Im$E$. Here ${i^\prime}$ represents all possible combinations of occupied states with positive imaginary parts (Im[$E$]$>0$).} \label{Hatano-figure-1}
\end{figure*}

\begin{figure*}
\includegraphics[scale=0.35]{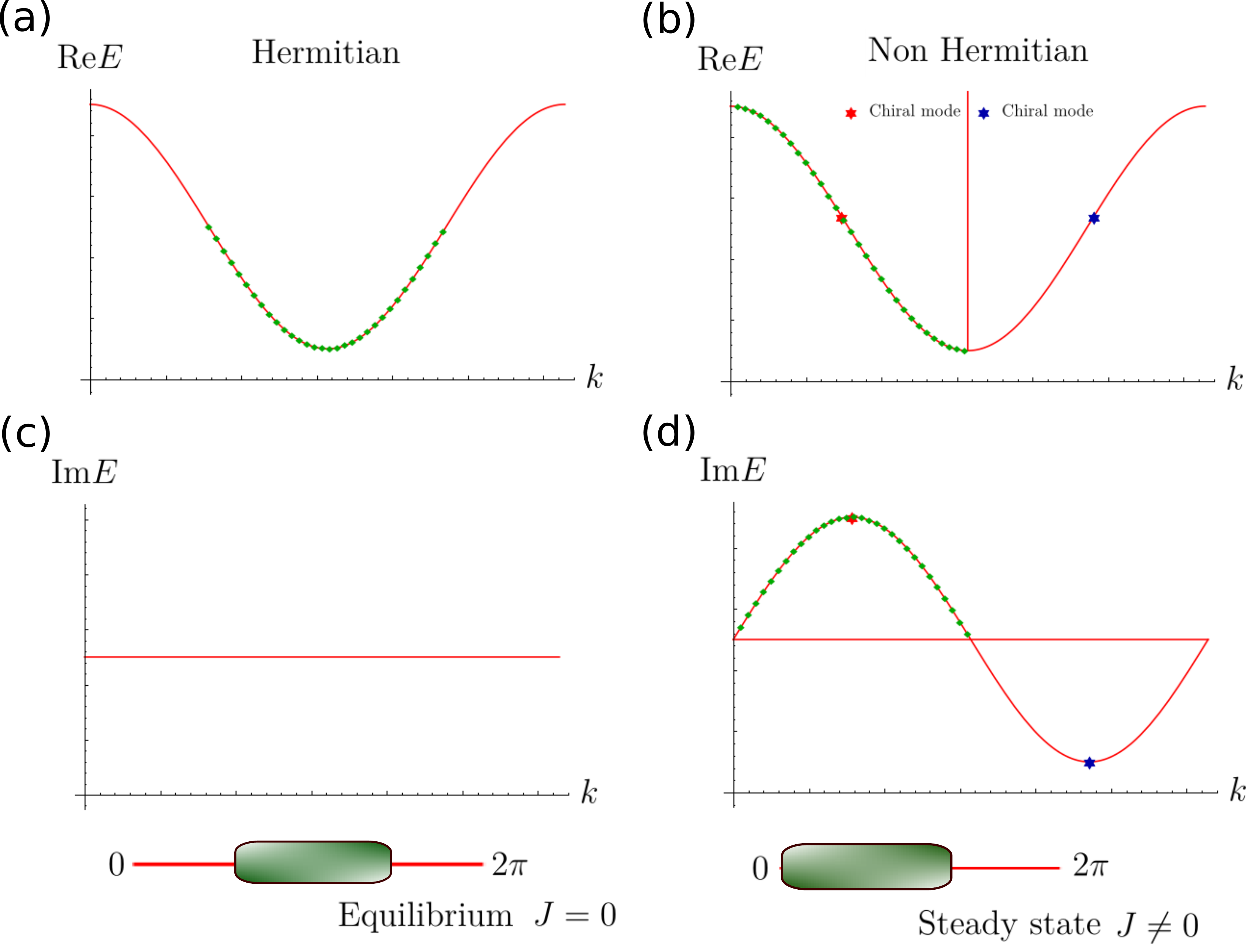}
  \caption{\textbf{Steady state particle filling and notion of NESS in the Hatano-Nelson model.} We illustrate the notion of many-body ‘steady states’ that arise in the non-interacting Hatano-Nelson model at a finite density of fermions (half-filling). (a) At the Hermitian limit, the half-filling of fermions is shown by green dots. The current contributions from various single particle states gets cancelled by each other, leading to an equilibrium state with zero current. (b) The notion of steady states in a non-Hermitian regime changes the particle filling. The system stabilizes a steady state with a Galilean-boosted Fermi sea. Consequently, the system drives a finite current. Here, the red and blue stars represent the chiral modes discussed in Section~\ref{models}A.}  \label{Hatano-figure-2}
\end{figure*}

Quantum many-body phases lead to novel manifestations beyond single particle physics, where the collective behavior of a large number of constituents offers several exotic phases of matter. Recently, there has been growing interest in combining the ideas of open quantum many-body phenomena and non-Hermitian topological phases~\cite{lee2020many}. The cross-fertilization between these two fields has led to new insights into the non-Hermitian phases. The construction of many-body states for non-Hermitian systems becomes particularly interesting since the ramified symmetry classes give rise to unique topological phases corresponding to the gapped single particle spectrum in the non-Hermitian regime. However, identifying the ground state and its low-energy excitations arising from the conventional real energy minimization, like in a Hermitian system, is now tricky in non-Hermitian systems due to its complex eigenspectra. The imaginary part of the energy introduces a certain lifetime corresponding to each single particle eigenstate. Consequently, the direct relation between the single particle eigenstates and many body states is missing, since the system may not attain an equilibrium state but rather a non-equilibrium steady state (NESS) at late times. NESSs become particularly important in the context of the driven open systems connected to leads~\cite{mahajan2016entanglement,eisler2014area,panda2020entanglement}. Recent studies show that the effective non-Hermitian Hamiltonian derived from such an open system is expected to attain a unique current carrying NESS as a response of the system to the external gauge field~\cite{panda2020entanglement}. The system reaches the NESS through the non-unitary evolution of the initial eigenstate, and the state at time $t$ can be written as~\cite{panda2020entanglement}

\begin{equation}
    |\psi (t) \rangle = \dfrac{\sum_n e^{-i \varepsilon_n t + \Lambda_n t} \langle n_L|\psi\rangle|n_R\rangle}{||\sum_n e^{-i \varepsilon_n t + \Lambda_n t} \langle n_L|\psi\rangle|n_R\rangle||},
\end{equation}

where  $\langle n_L|$ and $|n_R\rangle$ correspond to the left and right eigenvectors of the Hamiltonian with eigenvalue $E_n=\varepsilon_n+ i \Lambda_n$ and $||....||$ represents the norm of a vector. The NESS $(|\psi (t \rightarrow \infty) \rangle)$, leading to the maximum imaginary part of the eigenvalues, is given by~\cite{panda2020entanglement}

\begin{equation}
    |\psi (t \rightarrow \infty) \rangle =\dfrac{\langle s_L| \psi\rangle}{|\langle s_L| \psi\rangle|} \dfrac{|s_R\rangle}{||  |s_R\rangle||},
\end{equation}

where $\langle s_L |$ ($\langle s_R |$) denotes the eigenvector with the maximum imaginary part of the energy eigenvalue $E_n$ (see Fig.~\ref{Hatano-figure-1}). In Ref.~\cite{silberstein2020berry}, the authors analyze the wave-packet dynamics in a system governed by a non-Hermitian Hamiltonian based on a semiclassical treatment, and show that the state eventually leads to NESS while reframing the notion of the geometric phase. These recent studies on NESS redefine the idea of particle filling, which eventually gives rise to a many-body steady state at long times~\cite{lieu2019non}. Interestingly, the NESS has a one-to-one correspondence with non-Hermitian spectral topology. Their  intricate interplay has been recently discussed in Refs.~\cite{zhang2020correspondence,banerjee2022chiral}. The NESS corresponds to many-body steady states consisting of single-particle eigenstates with minimum real energy and maximum imaginary component~\cite{lieu2019non,banerjee2022chiral}. As we present in Fig.~\ref{Hatano-figure-2}, the occupied bands in the non-interacting Hatano-Nelson model (as discussed in Section Section~\ref{models}A) drive a finite current for the NESS, unlike any Hermitian system where current contributions from various single particle states cancel pairwise, respecting the Bloch theorem and leading to an equilibrium state with zero current. We refer the readers to Ref.~\cite{lieu2019non} for a conceptual realization of NESS. The author discusses the occurrence of symmetry-protected degenerate steady states in the topological phase of non-Hermitian Majorana modes, which arise in the effective Hamiltonian description of topological superconductors connected to a Markovian bath.

Along with these interesting studies on the non-Hermitian non-interacting systems, the theoretical understanding of NESS and their non-unitary dynamics in the presence of many-body interactions provides a promising research avenue for discovering qualitatively new phenomena unique to non-Hermitian systems. For instance, Kawabata \emph{et al.} demonstrated that non-Hermitian topological phases can survive in one-dimension even in the presence of many-body interactions~\cite{kawabata2022many}. Using the interacting Hatano-Nelson model, they further formulated a many-body topological invariant to illustrate the effect of such interactions on the NHSE. A recent study by Zhang \emph{et al.} shows that the Hatano-Nelson model in the presence of repulsive nearest-neighbour interactions exhibits two distinct phase transitions with increasing interaction strength~\cite{zhang2022symmetry}. The first transition was marked by EPs and characterized by first-order symmetry-breaking transitions to the charge density wave regime. In contrast, the second phase transition occurs at a critical interaction strength commensurate with the system size. The interplay between many-body localization phase transitions and associated quantum dynamics has become another fascinating aspect of current research considering the interplay between topology, interaction, quasiperiodicity and non-Hermiticity~\cite{zhai2020many,heussen2021extracting,hamazaki2019non,suthar2022non,mak2023statics,tang2021localization,hamazaki2022lindbladian}. Recently, Yoshida \emph{et al.} studied the reduction phenomena of the correlated gapped systems describing the fragility of exceptional topology under interactions~\cite{yoshida2022reduction,yoshida2023fate}. Classification of many-body quantum phases in accordance with their entanglement properties has become a major theme in quantum physics~\cite{altman2015universal,abanin2019colloquium,altman2015universal}. Recently, the ideas of the entanglement spectrum and associated dynamics have been extended in the non-Hermitian regime to detect the non-Hermitian topology as well as to characterize the many-body phases~\cite{orito2022unusual,turkeshi2022entanglement}. For instance, Herviou \emph{et al.} discussed the direct mapping between symmetries of the entanglement spectrum and that of the non-Hermitian Hamiltonian of a gapped system while considering a many-body state~\cite{herviou2019entanglement}. Furthermore, the entanglement entropy can characterize the many-body properties corresponding to the four non-equilibrium phases of the non-Hermitian SSH model as we discussed in Section~\ref{models}B. This approach can provide a theoretical understanding of the various phases and their transitions while uncovering the rich interplay of non-equilibrium many-body physics and topology~\cite{banerjee2022chiral}. As reported in recent studies, entanglement entropy and entanglement dynamics can act as an important diagnostic tools which reveal the topological signatures of a non-Hermitian many-body systems~\cite{kawabata2022entanglement,turkeshi2023entanglement,sayyad2021entanglement,lee2014entanglement}.

Another recent trend in this field has been the study of non-Hermitian random matrix theory (NH-RMT) in the context of open quantum many-body systems. The connection between NH-RMT and many-body open systems is particularly important in the study of quantum chaos. Non-Hermitian random matrices are an important tool for modelling the chaotic dynamics of many-body open systems, as they capture the complex interactions between the system and its environment. Recent studies reveal that by analyzing the statistical properties of the eigenvalues and eigenvectors of these matrices, one can gain deep insights into the chaotic behaviour of many-body open systems~\cite{garcia2022symmetry,hamazaki2020universality,li2021spectral,li2021spectral,neri2012spectra,xiao2022level}. Another intriguing aspect of NH-RMT in many-body open systems is the study of quantum systems that undergo Anderson transitions. In this context, non-Hermitian random matrices could prove to be very useful to model the effect of disorder and interaction with the environment on the energy spectrum of the system and to detect the critical behaviour at the transition points~\cite{goetschy2011non,molinari2009non,kawabata2021nonunitary}. As our understanding of these areas of physics continues to evolve, the connection between non-Hermitian topology and many-body open systems will likely continue to play an important role in this field.

\section{Physical Platforms and Experimental Advances}

We now present a brief survey of realizations of non-Hermitian Hamiltonians in various physical platforms featuring non-Hermitian topology, which in turn offers intriguing applications and novel functionalities in non-conserving settings. Dissipation and losses are expected in most of the physical systems, however, tuning dissipation in a controllable manner leads to wide variety of exotic phenomena with potential applications~\cite{bergholtz2021exceptional,ashida2020non}. Along with delicate theoretical beauty, the application of non-Hermitian topology can be found in both classical systems, including optical setups with gain and loss~\cite{midya2018non,parto2021non}, electric circuits~\cite{gupta2021non,wu2022non}, and quantum systems such as electronic transport setups at material junctions~\cite{bergholtz2019non,park2018quantum}, and dissipative cold-atom experiments~\cite{xu2017weyl,li2019observation,li2020topological,liang2022observation,li2022non}. The light-matter topological insulator with non-Hermitian topology in photonic crystals holds great promise both for fundamental discoveries and for opening the door to exciting applications in optoelectronics, lasing, as well as transport~\cite{feng2017non,el2019dawn,longhi2015robust,longhi2018parity,wang2021topological,de2022non}.

Recently, the intimate interplay of non-Hermiticity and topological photonics led to the realization of Weyl exceptional rings in an evanescently coupled bipartite optical waveguide array by introducing breaks between waveguides~\cite{cerjan2019experimental}. Combining the ideas of circular modulation of waveguides and insertion of breaks to achieve efficient gain and loss, one can tune non-Hermitian topological semimetals in different dimensions~\cite{banerjee2020controlling}. For an in-depth account of non-Hermitian topological photonics, we refer to the review by Parto \textit{et al.}, which features various photonic non-Hermitian platforms, EP dynamics, and recent advancements in these growing fields~\cite{parto2021non}. Mechanical systems such as acoustic materials and metamaterials provide another intriguing prospect for the experimental study of non-Hermitian topological phases. Recently, in the pioneering work of Ghatak \textit{et al.}~\cite{ghatak2020observation}, non-Hermitian topology of the non-reciprocal SSH model has been realized in terms of the dynamical matrix in robotic metamaterials consisting of mechanical rotors, control systems, and springs. These systems manifest non-reciprocal hopping through non-uniform strain in the springs connected to two adjacent rotors. Non-Hermitian topology delineating exceptional structure and concomitant phase transitions in mechanical metamaterials have also been reported in Refs.~\cite{zhou2020non,scheibner2020non,yoshida2019exceptional,brandenbourger2019non}. Interestingly, non-Hermitian chiral active matter systems have also recently emerged as a promising platform to experimentally realize the exceptional edge modes proposed by Sone \textit{et al.}~\cite{sone2020exceptional}.

Electric circuits present another classical platform which enables realization of non-Hermitian topology. In the classical topolectrical circuit framework, the circuit components, such as periodic arrays of capacitors and inductors, act as Hermitian elements, whereas non-Hermitian effects can be incorporated through dissipative resistance in such arrays. Such topoelectric setups have been shown to simulate diverse phases, and can also be used to readily model various non-Hermitian topological phases~\cite{albert2015topological,ningyuan2015time,imhof2018topolectrical,zhang2020non2,luo2018nodal,ezawa2019non,ezawa2019non2}. For instance, the circuit realization of the non-Hermitian SSH model and the associated skin effect has been reported by Helbig \textit{et al.}~\cite{helbig2020generalized} and Hofmann \textit{et al.}~\cite{hofmann2020reciprocal}. Thus, these topoelectrical circuits serve as a tunable classical platform for theoretical modeling and experimental realization of non-Hermitian topological band structures.

Several ingenious experimental setups have been suggested and realized to engineer EPs in diverse physical platforms. Most notable among them have been in magnetic multi-layers~\cite{yu2020higher}, waveguides~\cite{zhong2016parametric,zhang2019dynamically}, topoelectric circuits~\cite{schindler2011experimental,stegmaier2021topological,xiao2019enhanced}, photonic lattices~\cite{feng2017non,parto2021non}, and topological insulator-ferromagnet junctions~\cite{bergholtz2019non}. Recent theoretical proposals and experimental demonstrations show that a bulk Fermi arc arises from the topological protection of EPs in an open system of photonic crystal slabs~\cite{zhou2018observation}. Interestingly, the bulk Fermi arc is fundamentally different from the conventional surface Fermi arcs that emerge from the two-dimensional projection of Weyl points in three-dimensional Hermitian systems~\cite{armitage2018weyl}. Furthermore, it leads to fractional topological charge, and the unusual topological band-switching properties across the Fermi arc have also been very recently observed~\cite{zhou2018observation,su2021direct,tang2020exceptional}. In addition, the critical behaviour around EPs in the complex parameter space often leads to dramatic effects in a wide range of fields~\cite{ashida2020non,bergholtz2021exceptional}, and the theory of EPs has become one of the most exciting emergent front at the crossroads of optics and photonics, acoustics, quantum physics and atomic physics~\cite{zhang2016observation,ozawa2019topological,yao2018edge,feng2014single,shi2016accessing,midya2018non}. Its explorations have led to many intriguing proposals, such as uni-directional sensitivity~\cite{lin2011unidirectional,peng2014parity}, band merging~\cite{zhen2015spawning,makris2008beam}, laser mode selectivity~\cite{feng2014single,hodaei2014parity}, optomechanical energy transfer~\cite{xu2016topological}, and interesting quantum dynamics of exciton-polaritons~\cite{gao2018chiral,gao2015observation}, to name just a few. The EPs in $PT$-symmetric systems emerge naturally in open quantum platforms owing to a balanced gain and loss interaction between the system and the environment. They result in a wide range of applications, such as single mode coherent lasing~\cite{feng2014single}, negative refraction~\cite{fleury2014negative}, enhancement of sensing~\cite{hodaei2017enhanced,wiersig2020prospects} and unidirectional cloaking~\cite{sounas2015unidirectional}, due to the unique kind of eigenstructures in $PT$ broken and unbroken parameter regimes~\cite{feng2017non,longhi2018parity,zyablovsky2014pt,miri2019exceptional}.

An exciting recent experimental development was made in realizing the NHSE in a photonic system by Weidemann \textit{et al.}~\cite{weidemann2020topological}, where non-Hermiticity was invoked by tailoring anisotropy in the nearest-neighbour coupling in an optical fibre. This anisotropy was controlled by amplitude and phase modulators. The authors discovered that when an interface is introduced in the system, all the eigenstates travel towards the interface and localize there, which essentially acts like a ``topological funnel" for the light field and demonstrates the NHSE. Further, Zhang \textit{et al.}~\cite{zhang2021observation} demonstrated the occurrence of a higher-order NHSE in an acoustic topological insulator where opposite spins can be localized at opposite corners of the sample by controlling the nature and strength of the non-Hermiticity. Other aspects of NHSE have been recently summarized in Ref.~\cite{zhang2022review}. There exist a few reports of tuning transport in non-Hermitian topological systems. For instance, topologically resilient light can be directed and guided along any arbitrary route thanks to non-Hermitian control in photonic topological insulators~\cite{zhao2019non}. Zhao \textit{et al.} used optical pumping to introduce non-Hermiticity into a coupled microring resonator system, which generated dispersed gain (via external pumping) and loss (intrinsic material loss), and the topological pathways for light were created due to local non-Hermiticity at the boundary of the gain and loss domain. This customizable non-Hermitian topological navigation of light opens up possibilities for high-density data processing in integrated photonic circuitry.

\section{Overview and Outlook}

In summary, we presented key concepts underlying the physics of non-Hermitian topological phases. By means of concrete, paradigmatic models -- Hatano-Nelson, non-Hermitian Su-Schrieffer-Heeger, and non-Hermitian Chern insulator -- we illustrated the important features of non-Hermitian topological systems. We summarized the non-Hermitian symmetry classes and discussed the notion of complex energy gaps. We highlighted the non-Hermitian skin effect and the role of generalized Brillouin zone in restoring the bulk-boundary correspondence. We examined the role of disorder and the linear response of such non-Hermitian phases. We also surveyed the rapidly growing experimental advances made in the field.

We would like to end by highlighting here some possible directions which, in our view, may be promising for future explorations. The role of interactions in non-Hermitian topological phases remains to be fully unravelled. While there are a few initial promising studies~\cite{kawabata2022many,hyart2022non,chen2022topological,suthar2022non,ghosh2022spectral,faugno2022interaction,yoshida2022fate,yoshida2022reduction,yoshida2018non,crippa2021fourth,schafer2022symmetry,shen2021non,zhang2022symmetry,sarkar2021study,kumar2021emergence}, a complete picture is lacking at present. Another important aspect is to understand the time dynamics and steady state nature of such non-Hermitian topological phases and their phase transitions~\cite{longhi2019probing,longhi2022non,li2022engineering,zhou2019dynamical,moca2021universal,silberstein2020berry,para2021probing,longhi2022self,rahul2022topological}. Preliminary investigations have revealed the presence of intriguing non-equilibrium steady state phases that may be characterized by the entanglement properties~\cite{banerjee2022chiral}. The role of dimensionality in this regard may also prove to be important. An open question is the connection between the Liouvillian frameworks, which are well-established for open quantum systems~\cite{breuer2002theory}, and effective non-Hermitian Hamiltonian approaches, which are predominantly used for non-Hermitian topological phases. We note that there are a few recent forays into exploring this connection~\cite{song2019non,haga2021liouvillian,yang2022liouvillian,longhi2020unraveling,liu2020helical,gong2022bounds,roccati2023hermitian}, although a full understanding is as yet out of reach. Machine learning and artificial intelligence tools are growing in importance in physical sciences~\cite{carleo2019machine}. Very recently, there have been a handful of studies using both supervised and unsupervised machine learning approaches to characterize non-Hermitian topological phases~\cite{narayan2021machine,zhang2021machine,yu2021unsupervised,araki2021machine,yu2021experimental}. Several connections to modern mathematics are also worth exploring in the context of non-Hermitian systems~\cite{wang2022amoeba,banerjee2023tropical}. Use of such methods to discover and design non-Hermitian systems presents a tantalizing prospect. In conclusion, we are optimistic that many interesting aspects of non-Hermitian topological phases remain to be discovered, and we hope that the present review can help crystallize some of them.

\section*{Acknowledgments}

We are grateful to Adhip Agarwala, Arka Bandyopadhyay, Amartya Bose, Nisarg Chadha, Debashree Chowdhury, Tanmoy Das, Suraj Hegde, Rimika Jaiswal, Madhusudan Manjunath, Tobias Meng, Subroto Mukerjee, Brajesh Narayan, Afsar Reja, and Vijay Shenoy for discussions and collaborations. A. B., R. S., and S. D. are supported by the Prime Minister's Research Fellowship (PMRF). A. N. acknowledges support from the startup grant of the Indian Institute of Science (SG/MHRD-19-0001) and DST-SERB (project number SRG/2020/000153).

\bibliography{bibliography.bib}

\end{document}